\documentclass[aps,prl,reprint,superscriptaddress,floatfix]{revtex4-2}
\usepackage{graphicx}
\usepackage{dcolumn}
\usepackage{bm}
\usepackage{amsmath}
\usepackage{overpic}
\usepackage{subfigure}
\usepackage{ragged2e}
\usepackage{amssymb}

\usepackage[colorlinks=true,citecolor=blue,urlcolor=blue]{hyperref}

\begin{document}

\title{Topological States Enabled by Non-local Nonlinearity in Synthetic Dimensions}

\author{Chong-Xiao Chen}
\affiliation{Anhui Province Key Laboratory of Quantum Network, University of Science and Technology of China, Hefei 230026, China}
\affiliation{CAS Center For Excellence in Quantum Information and Quantum Physics, University of Science and Technology of China, Hefei 230026, China}

\author{Zheng-Wei Zhou}\email{zwzhou@ustc.edu.cn}
\affiliation{Anhui Province Key Laboratory of Quantum Network, University of Science and Technology of China, Hefei 230026, China}
\affiliation{CAS Center For Excellence in Quantum Information and Quantum Physics, University of Science and Technology of China, Hefei 230026, China}
\affiliation{Hefei National Laboratory, University of Science and Technology of China, Hefei 230088, China}
\affiliation{Anhui Center for Fundamental Sciences in Theoretical Physics, University of Science and Technology of China, Hefei, 230026, China}

\author{Han Pu}
\email{hpu@rice.edu}
\affiliation{Department of Physics and Astronomy, and Smalley-Curl Institute, Rice University, Houston, Texas 77251-1892, USA}

\author{Xi-Wang Luo}\email{luoxw@ustc.edu.cn}
\affiliation{Anhui Province Key Laboratory of Quantum Network, University of Science and Technology of China, Hefei 230026, China}
\affiliation{CAS Center For Excellence in Quantum Information and Quantum Physics, University of Science and Technology of China, Hefei 230026, China}
\affiliation{Hefei National Laboratory, University of Science and Technology of China, Hefei 230088, China}
\affiliation{Anhui Center for Fundamental Sciences in Theoretical Physics, University of Science and Technology of China, Hefei, 230026, China}

\date{\today}

\begin{abstract}
The interplay between topology and nonlinearity represents a central challenge in modern physics. 
Here, we investigate this interplay by considering a synthetic Su-Schrieffer-Heeger lattice with all-to-all nonlocal interactions. We find that the distinctive nonlinearity maintains an effective chiral symmetry and leads to a quantized nonlinear winding and Berry phase, as corroborated by the developed Bogoliubov nonlinear adiabatic theory. 
Increasing nonlinearity drives a sequence of topological transitions signaled by the appearance of characteristic swallowtail band structures at intermediate interaction strengths and band swapping in the strong nonlinear regime.
The band swapping results in quantized fractional windings and double-period Bloch oscillations that are closely related to discrete time crystals. Remarkably, even starting from a topologically trivial linear system, nonlocal nonlinearity can induce an emergent topological phase with fractional windings.
Experimentally, our model can be realized using photons in a degenerate optical cavity with Rydberg-mediated interactions. Our results establish a rigorous framework and pave the way for exploring nonlinear topological phenomena and their applications in synthetic quantum platforms.
\end{abstract}

\maketitle
\textit{\textcolor{blue}{Introduction}}---Topological phases, characterized by bulk topological invariants and disorder-robust edge states~\cite{Qi2011TopologicalInsulators,Hasan2010ColloquiumTopological}, have been extensively explored beyond solid-state materials in platforms including photonics~\cite{Ozawa2019TopologicalPhotonics,Skirlo2015ExperimentalObservation,Hafezi2011RobustOptical,Rechtsman2013PhotonicFloquet,Kraus2012TopologicalStates,Song2025ArtificialGauge}, cold atoms~\cite{Goldman2016TopologicalQuantum,Cooper2019TopologicalBands,Gross2017QuantumSimulations}, acoustics~\cite{Ma2019TopologicalPhases,Xue2022TopologicalAcoustics,Zhang2018TopologicalSound}, and electrical circuits~\cite{Ningyuan2015TimeSiteresolved,Hafezi2013ImagingTopological,Lee2018TopolectricalCircuits}. While the linear topological band theory is well established, a major ongoing effort seeks to connect topology with nonlinearity~\cite{Sone2024NonlinearityinducedTopologicala,Hadad2016SelfinducedTopological,Pal2018AmplitudedependentTopological,Zhou2022TopologicalInvarianta,Tuloup2020NonlinearityInduceda,Liu2010BerryPhasea,Sone2025TransitionTopologicala,Wu2005GeometricPhasea,Liu2018NonlinearAdiabatic,Hu2025NonlinearGeometric}. Nonlinear effects are intrinsic to many platforms, such as Bose-Einstein condensates in cold atoms~\cite{Griffin1996BoseeinsteinCondensation,Dalfovo1999TheoryBoseeinstein} or Kerr nonlinearity in optics~\cite{Smirnova2020NonlinearTopological,Stolen1973OpticalKerr,Gross1961StructureQuantized}, whose dynamics is characterized by the mean-field Gross–Pitaevskii equations and Bogoliubov quantum excitations. Recent research has uncovered phenomena such as nonlinearity-induced topological phase transitions~\cite{Sone2024NonlinearityinducedTopologicala} and amplitude-dependent edge states across one and two dimensions~\cite{Hadad2016SelfinducedTopological,Pal2018AmplitudedependentTopological,Zhou2022TopologicalInvarianta}, stimulating the development of novel nonlinear topological invariants. Studies have also revealed unique interplay between topological edge modes and non-equilibrium phenomena, including solitons~\cite{Smirnova2019TopologicalEdge} and synchronization~\cite{Wachtler2020DissipativeNonequilibrium}. Up to now, the predominant focus has remained on real-space lattices with local nonlinear interactions.

In this work, we investigate the interplay between nonlocal nonlinearity and topology using a synthetic Su-Schrieffer-Heeger (SSH) lattice with all-to-all density and exchange interactions. By developing a Bogoliubov nonlinear adiabatic theory, we demonstrate that the system hosts a quantized nonlinear winding number and Berry phase, protected by an effective chiral symmetry under the nonlocal nonlinearity. We find that increasing the nonlinearity drives a sequence of topological phase transitions marked by the emergence of characteristic swallowtail band structures at intermediate nonlinearities and band swapping at strong nonlinearities, along with transitions of nonlinear edge states. The band swapping across the Brillouin zone (BZ) leads to fractional windings and Bloch oscillations with period doubling that are closely linked to discrete time crystals.
Remarkably, fractional windings emerge for strong nonlocal nonlinearity regardless of the underlying linear topology, which highlights a key difference from systems with local interactions.
We propose to implement the synthetic lattice using the orbital angular momentum (OAM) of photons within a degenerate cavity~\cite{Luo2015QuantumSimulationa,Luo2018TopologicalPhotonic} with all-to-all interactions induced by coupling photons to a Rydberg atomic ensemble~\cite{vsibalic2018rydberg,Georgakopoulos2018TheoryInteractinga,Pritchard2010CooperativeAtomlight,clark2020observation}. Our results can be extended to a broader class of synthetic lattice models and retain their validity even under realistic imperfections.

\textit{\textcolor{blue}{The model}}---We consider a synthetic SSH lattice model with single particle Hamiltonian 
\begin{equation}
    \mathcal{H}_0={\sum_{l}}\left[ J + (-1)^l \delta J \right] \hat{c}_{l+1}^\dagger \hat{c}_l + h.c.,
\end{equation}
where $\hat{c}_l^\dagger$ is the bosonic particle creation operator for synthetic site $l$, with intra- and inter-cell tunneling rates $J_{1,2}=J\mp\delta J$. 
We consider an all-to-all nonlocal interaction in the synthetic space 
\begin{equation}
    \mathcal{H}_\text{int}=-\frac{g}{2\pi}\sum_{l_1,l_2,l_3,l_4} \delta_{l_1+l_2,l_3+l_4} \hat{c}_{l_1}^\dagger  \hat{c}_{l_2}^\dagger \hat{c}_{l_3} \hat{c}_{l_4},
    \label{eq:Hint}
\end{equation}
with $g$ the interaction strength. Note that although such all-to-all interactions are typically not present in real materials, they arise quite naturally in synthetic dimensions. An experimental realization will be discussed later. By representing the sites using unit-cell and sublattice indices $\hat{c}_{2n-1}\rightarrow \hat{a}_n $ and $\hat{c}_{2n}\rightarrow \hat{b}_n$, the total Hamiltonian $\mathcal{H}_\text{tot}=\mathcal{H}_0+\mathcal{H}_\text{int}$ in the Bloch momentum space is 
\begin{equation}
    \mathcal{H}_\text{tot}=
    \int dk \,\hat{\psi}_k^\dagger \begin{pmatrix}
        -g\hat{n}_k&\hat{h}^\dagger\\
        \hat{h}&-g\hat{n}_k
    \end{pmatrix} \hat{\psi}_k,
    \label{eq:Ham}
\end{equation}
where $\hat{\psi}_k=[\hat{a}_k,\hat{b}_k]^T$ with $\hat{a}_k=\frac{1}{\sqrt{2\pi}}\sum_n \hat{a}_n e^{-ink}$ (similar for $\hat{b}_k$), $\hat{h}=J_1+J_2e^{ik}-g\hat{a}^\dagger_k\hat{b}_k-g\hat{a}_k \hat{b}_k^\dagger e^{ik}$, and $\hat{n}_k=\hat{a}_k^\dagger \hat{a}_k+\hat{b}_k^\dagger \hat{b}_k$. The Heisenberg equation of motion for $\hat{\psi}_k$ can be derived as~\cite{SUPP}
\begin{equation}
    i\partial_t \hat{\psi}_k = [\hat{\psi}_k,\mathcal{H}_\text{tot}]=\hat{H}_\text{eff}(k)\hat{\psi}_k=\begin{pmatrix}
        -2g\hat{n}_k& \hat{h}^\dagger_\text{eff}\\
        \hat{h}_\text{eff}&-2g\hat{n}_k
    \end{pmatrix}\hat{\psi}_k,
\end{equation}
with $\hat{h}_\text{eff}=J_1+J_2e^{ik}-2g\hat{a}_k^\dagger\hat{b}_k-2g\hat{a}_k \hat{b}^\dagger_k e^{ik}$. Note that the phase factor $e^{ik}$, a key determining factor of winding number, enters directly into the nonlinear term.

A crucial feature of our synthetic lattice is that its unique long-range interaction is naturally diagonal in Bloch momentum space, making $k$ a good quantum number and eliminating the need for additional assumptions required in real-space lattices. Although this interaction breaks the chiral symmetry of the full Hamiltonian, the dynamical Hamiltonian $\hat{H}_\text{eff}(k)$, satisfying $\sigma_z \hat{H}_\text{eff}(k) \sigma_z=-\hat{H}_\text{eff}(k)-4g{\hat n_k}$, effectively recovers chiral symmetry when considering a fixed {particle number $\langle\hat n_k\rangle=\rho_0$ for all modes $k$} and neglecting the constant energy shift.

\textit{\textcolor{blue}{Nonlinear topological invariant}}---The mean-field stationary solution can be obtained by treating the field operators as $c$-numbers (that represent the coherent-state wavefunctions)~\cite{Griffin1996BoseeinsteinCondensation,Dalfovo1999TheoryBoseeinstein,Smirnova2020NonlinearTopological,Stolen1973OpticalKerr,Gross1961StructureQuantized} and solving the nonlinear eigenequation $H_\text{eff}(k) \psi_{m,k}=E_{m,k} \psi_{m,k}$~\cite{SUPP}. 
Due to the effective chiral symmetry, the solutions take the form $\psi_{m,k}=\sqrt{\rho_0}{\chi_{m,k}}$ with ${\chi_{m,k}}=\frac{1}{\sqrt{2}}[1, e^{i\varphi_{m,k}}]^T$, where $\varphi_{m,k}$ can be solved self-consistently through $\arg[\pm h_\text{eff}]=\varphi_{m,k}$, with $h_\text{eff}=J_1+J_2e^{ik}- Ue^{i\varphi_{m,k}}- Ue^{-i\varphi_{m,k}} e^{ik}$, $U=g\rho_0$, and ``$\pm$" representing two bands. The eigenenergies are $E_{m,k}=\pm|h_\text{eff}(k)|$ where we notice that $|h_\text{eff}|$, due to interaction, is state-dependent and takes different values for different bands. Thanks to the effective chiral symmetry, we can define a nonlinear winding number
\begin{equation}
 W_m=\frac{1}{2\pi}\int_0^{2\pi} dk \,\partial_k\operatorname{arg}[h_\text{eff}(\psi_{m,k})]   
 \label{eq:winding}
\end{equation}
to characterize the topology of the nonlinear bulk modes. 

Alternatively, we can define a nonlinear Berry phase by evolving the Bloch momentum adiabatically across the BZ. The equivalence between winding number and Berry phase is not guaranteed in nonlinear systems due to the state dependence of $H_\text{eff}$, and the excitation during evolution introduces an energy shift that may accumulate a net geometric phase~\cite{Liu2018NonlinearAdiabatic,Hu2025NonlinearGeometric}. Here, we explicitly develop a general Bogoliubov nonlinear adiabatic theory with slowly varying $k(t)=k(0)+\epsilon t$. We expand the wavefunction as $\psi_k(t)=e^{i\gamma_m(t)-i\int^t E_{m,k(t')}dt'}[\sqrt{\rho_0} {\chi_{m,k}}+\delta\psi_k(t)]$, with Bogoliubov excitation
\begin{equation}
    \delta\psi_k(t)= {\chi_{B,k}}  [u_k\alpha e^{-i\int^{t}\omega_{k(t')} dt'}+v_k^{*}\alpha^{*}e^{i\int^{t}\omega_{k(t')} dt'}], \nonumber
\end{equation}
where $(u_k,v_k)$ is the instantaneous Bogoliubov mode~\cite{Baillie2017CollectiveExcitationsa,pitaevskii2016bose} with frequency $\omega_k$, excitation amplitude $\alpha$, and instantaneous {state} ${\chi_{B,k}}$ orthogonal to ${\chi_{m,k}}$. To first order in $\epsilon$, {the nonlinear Berry phase $\gamma_m(t)$ satisfies}~\cite{SUPP}
\begin{equation}
    \frac{d}{dt}\gamma_m(t)=i\langle\chi_{m,k(t)}|\partial_t|\chi_{m,k(t)}\rangle-A_{\text{nl}}(t)\label{phase:m},
\end{equation}
{with gauge choice $\gamma_m(0)=0$ and $|\chi_{m,k}\rangle$ the Dirac notation for $\chi_{m,k}$}. {Besides the Berry connection associated with the nonlinear state, there is an additional} geometric contribution from the nonlinear dynamical phase, with $A_{\text{nl}}(t)=\langle\chi_{m,k}|\sqrt{\rho_0}H^{(1)}_\text{eff}(\alpha)|\chi_{m,k}\rangle$, $H^{(1)}_\text{eff}$ the first-order correction of $H_{\rm eff}$ which depends on $\alpha(t)={\sqrt{\rho_0}}(A_1u_k^*+A_1^*v_k^*)e^{i\int ^t \omega_{k(t')} dt'}$ and $A_1=i\langle\chi_{B,k}|\partial_t|\chi_{m,k}\rangle/\omega_k$ (the adiabatic condition requires $|A_1|\ll 1$)~\cite{SUPP}. Our theory provides a complete description of nonlinear adiabatic dynamics by incorporating the full spectrum of particle-hole excitations, in contrast to previous semi-classical approaches with only particles or based on simplified assumptions~\cite{Liu2010BerryPhasea,Tuloup2020NonlinearityInduceda,Pu2007AdiabaticCondition}. This yields a unified and more intuitive physical framework.

\begin{figure}[t]
    \centering
    \includegraphics[width=0.48\textwidth]{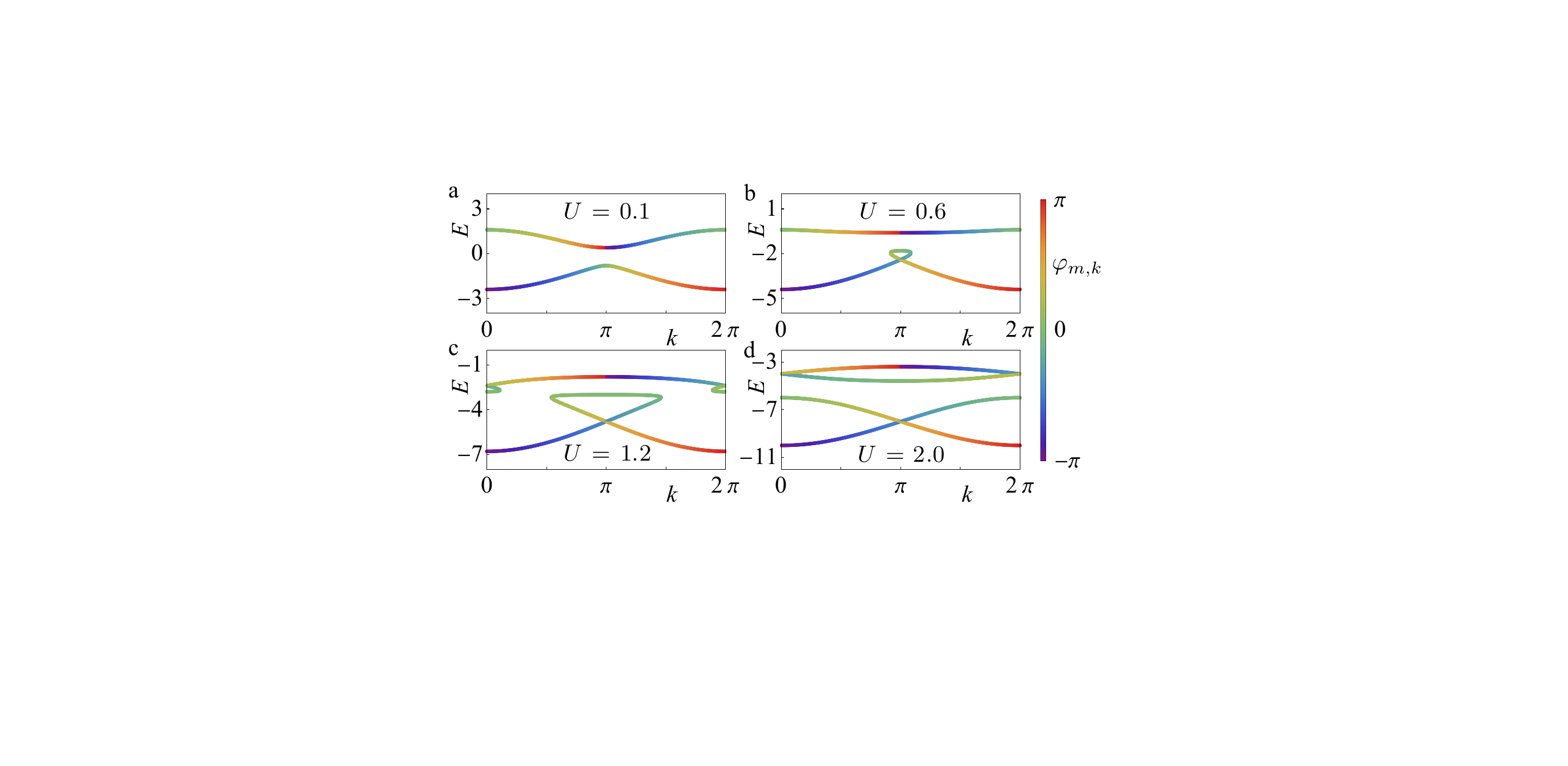}
	\caption{\justifying{\textbf{Nonlinear bands and winding properties.} With increasing nonlinearity: (a) Two well-defined bands. (b) Swallowtail emerges in the lower band. (c) Swallowtail forms in both bands. (d) Swallowtails connect into a four-band structure.  Color bar indicates the phase $\varphi_{m,k}$, whose integral yields the winding number. $J=1$ and $\delta J=0.3$ in all plots.}}\label{fig:energybands}
\end{figure}

$A_{\text{nl}}$ can be obtained through a straightforward but involved calculation. For the nonlinear synthetic lattice with effective chiral symmetry, we find $A_{\text{nl}}=0$~\cite{SUPP} and the nonlinear Berry phase along the BZ reduces to $\gamma_m{(t=\frac{2\pi}{\epsilon})}=\pi W_m$, with $W_m$ the quantized winding number. This direct quantization of $\gamma_m$ stems from the unique nonlocal nonlinearity in our system. This behavior presents a fundamental departure from real-space lattices with local interactions, which yield a non-quantized Berry phase even under the assumption of single-Bloch component~\cite{Liu2010BerryPhasea,Tuloup2020NonlinearityInduceda,Zhou2022TopologicalInvarianta}.

\begin{figure}[t]
    \centering
    \includegraphics[width=0.48\textwidth]{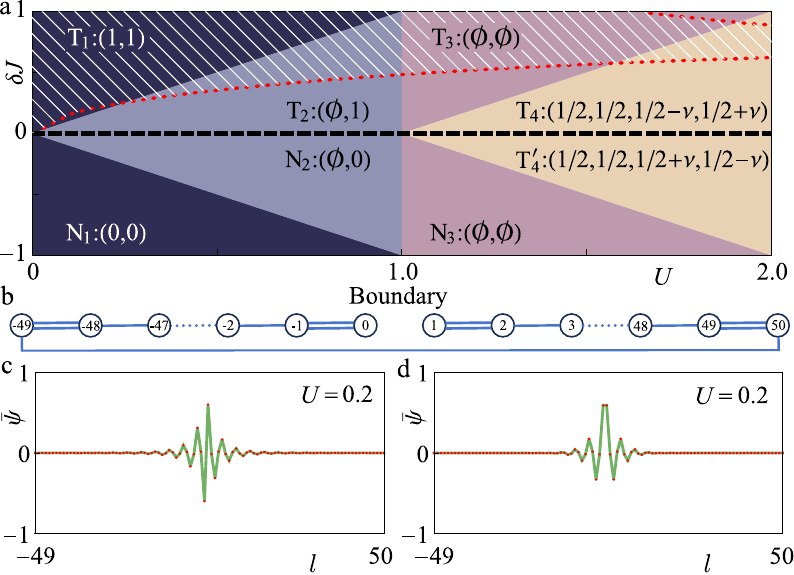}
	\caption{\justifying{\textbf{Phase diagram and edge states.} (a) The winding properties of different phases encoded by different colors, with $\varnothing$ denoting the ill-defined winding number. Phases $\text{T}_4$ and $\text{T}_4'$ have four well-defined bands (i.e., four winding numbers). (b) The synthetic lattice with open boundary at $l = 0$ and $l = 1$. (c, d) The anti-symmetric and symmetric edge states for the lattice geometry in (b), with $\delta J=0.3$, $U=gN_{\rm edge}=0.2$. $\bar\psi=\psi/\sqrt{N_{\rm edge}}$ is the normalized wave function. For $\delta J>0$, two edge-state solutions exist {in the shaded region, while only the antisymmetric solution persists in the unshaded region, as delineated by the red-dotted lines in panel (a).}
    }}\label{fig:phasediagram}
\end{figure}

\textcolor{blue}{\textit{Phase diagram}---}In Fig.~\ref{fig:energybands}, we plot the nonlinear band structures and winding properties for various nonlinear strengths. The corresponding phase diagram is shown in Fig.~\ref{fig:phasediagram}. When defining the $m$-th band and its winding $W_m$, we require $\psi_{m,k}$ to follow a smooth path as $k$ varies from $0$ to $2\pi$. 
We first focus on the region $\delta J>0$ where the linear Hamiltonian is topological. 
For weak nonlinearity $U<|\delta J|$ (phase $\text{T}_1$), we find two smooth nonlinear bands which are gapped and dynamically stable with finite Bogoliubov gap in the whole BZ. Therefore, both the winding number and Berry phase are well defined and quantized to $W_m=1$, {as inferred from the phase winding shown in Fig.~\ref{fig:energybands}a.}

As the nonlinearity increases to the region $U\in[|\delta J|,J]$ (phase $\text{T}_2$), the lower band develops a crossing at $k=\pi$ and forms a swallowtail structure, as shown in Fig.~\ref{fig:energybands}b. The upper branch of the emergent swallowtail loop is dynamically unstable  with complex Bogoliubov energy. The winding number and Berry phase for the lower band become ill-defined due to instability and discontinuity of the solution as a function of $k$. Further increasing the nonlinearity to the region $U\in[J,J+|\delta J|]$ (phase $\text{T}_3$) leads to the swallowtail structure in the upper band at $k=0$, as shown in Fig.~\ref{fig:energybands}c, which also features an unstable upper branch. Consequently, both bands lack a well-defined topological invariant, though the phase winding along a closed band trajectory (including the swallowtail loop) is still quantized to 1.
The swallowtail structure is generic in nonlinear systems, but usually appears in the parameter space~\cite{Karkuszewski2002MeanFielda,Mulansky2011ImpurityBoseeinstein,Wu2006CommutabilitySemiclassicala,Liu2003NonlinearEvolutiona} rather than the Bloch-momentum space as in our nonlinear synthetic lattices.

In the strong nonlinear regime $U>J+|\delta J|$ (phase $\text{T}_4$), the two swallowtail structures merge and subsequently open a gap, resulting in 4 well-defined bands within the BZ (with the third band dynamically unstable), as shown in Fig.~\ref{fig:energybands}d. The 4 bands are grouped into two sectors by an energy gap. We find that the winding numbers $W_{1,2}$ are fractionally quantized to 1/2, while $W_{3,4}=1/2\pm \nu$ are not quantized, where $\nu$ depends on $|\delta J|$ and $U$.
In the regime $\delta J<0$, the nonlinear band structures are similar but with different winding properties.  
Based on the analysis of band structure and winding number, we identify a total of eight phases, as summarized in the phase diagram of Fig.~\ref{fig:phasediagram}. 
In particular, an emergent topological phase $\text{T}_4'$ with $W_{1,2}=1/2$ arises for strong nonlinearities, where the topology is dominated by the phase factor of the nonlinear term.
Therefore, fractional windings constitute an intrinsic property of strong nonlocal nonlinearity {in Eq.~\eqref{eq:Hint}}, regardless of the underlying linear topology. This highlights a key difference from local-interacting systems, where no fractional winding occurs.

\begin{figure}[t]
    \centering
    \includegraphics[width=0.48\textwidth]{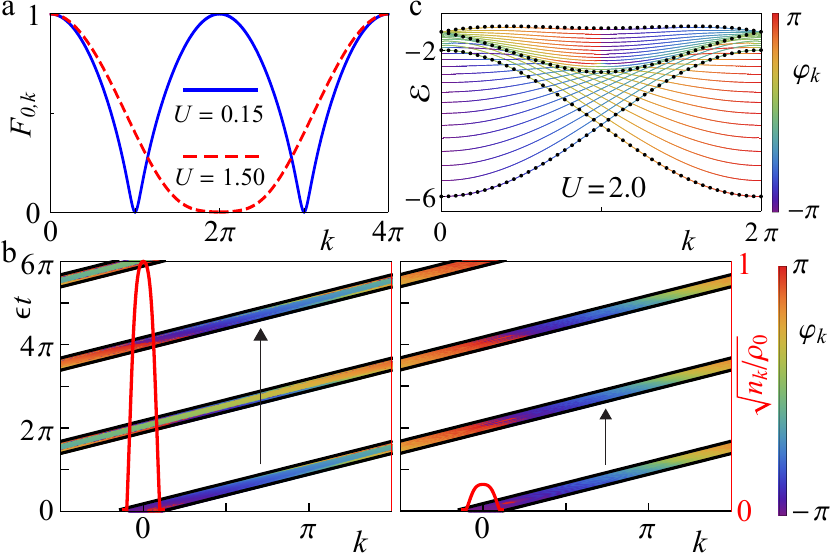}
	\caption{\justifying{\textbf{Nonlinearity induced period multiplexing.} (a) Overlap of low-energy eigenstates between $k\ne0$ and $k=0$. It recovers the initial value after $k$ traverses the BZ once (twice) for weak (strong) nonlinearity. (b) {Bloch oscillations} of two wave packets initialized in the instantaneous eigenstate at $k=0$. Color bar shows the phase evolution $\arg[a_k^*(t)b_k(t)]$ during propagation along $k(t)$ with $\epsilon=0.01$ and $g\rho_0={3.0}$. The large-amplitude wave packet with $g{n_k}>J+|\delta J|$ exhibits period-doubling, compared with weak-amplitude ordinary {Bloch oscillation} with $g{n_k}<|\delta J|$, as marked by the vertical arrows. (c) Full quantum bands (solid lines) with ${n_k}=27$ and $U=g{n_k}$ {($\forall k$)}, 
    comparing to the mean-field results (black dots). 
    $\mathcal{E}$ represents energy per particle.
    Color bar indicates $\varphi_k=\arg[\langle \hat{a}^\dagger_k \hat{b}_k\rangle]$. $J=1$, $\delta J=0.3$ in all plots.
  }}\label{fig:period}
\end{figure}

Omitting the energy shift $gn_k$ reduces the nonlinearity to a purely intersublattice coupling; thus, the linear edge state (localized on one sublattice) and the bulk-edge correspondence remain unaltered. In realistic settings, an ideal boundary is difficult to realize because nonlocal interactions inherently couple to synthetic sites beyond any single-particle boundary (see Fig.~\ref{fig:phasediagram}b). Moreover, the $k$-dependent density of the edge state prevents $gn_k$ from being omitted as a constant. Nevertheless, by solving the mean-field eigenstates in OAM-lattice space, we find two edge-state solutions for $\delta J>0$ in the small-$U$ regime, one symmetric and the other anti-symmetric, with only the anti-symmetric solution persisting in the large-$U$ regime~\cite{SUPP}, as shown in Figs.~\ref{fig:phasediagram}c and \ref{fig:phasediagram}d. Since for the symmetric edge state, nonlocal nonlinearity weakens the nearest-neighbor tunnelings $J_{1,2}$, which effectively amplify the long-range nonlinear coupling and delocalize the state. In contrast, $J_{1,2}$ is enhanced for the anti-symmetric edge state.
For $\delta J<0$, there are no topological edge solutions even for the emergent topological phase $\text{T}_4'$. Hence we conclude that although the band topology could be dominated by strong nonlinearity, the appearance of edge state is full determined by the staggered single-particle tunneling.

\textit{\textcolor{blue}{Fractional winding and period doubling}}---The strong nonlinear regime ($\text{T}_4$ and $\text{T}_4'$) features four well-defined bands, which undergo mutual exchanges as $k$ varies from $0$ to $2\pi$, leading to fractional winding of each band. The nonlinear eigenstates also exchange with each other as $k$ varies across the BZ, as clearly seen by examining the overlap of the eigenstates between momenta $0$ and $k$: $F_{0,k}=|\langle\chi_{1,0}| \chi_{m,k}\rangle|^2$. As shown in Fig.~\ref{fig:period}a, for weak nonlinearity ($U<|\delta J|$), $F_{0,k}$ recovers its initial value after $k$ traverses the BZ once. By contrast, for strong nonlinearity ($U>J+|\delta J|$), $k$ needs to traverse the BZ twice before $F_{0,k}$ recovers.
This period doubling constitutes a fundamental nonlinear effect different from single-particle systems with specific symmetries such as glide-reflection~\cite{Khan2024ObservationPerioddoublinga,Li2022BoseeinsteinCondensate,Zhang2017TwolegSuschriefferheeger,Holler2018TopologicalBloch} or non-Hermitian parity-time~\cite{PhysRevLett.116.133903,PhysRevLett.120.146402} symmetries. The underlying principle directly enables the generalization to period multiplexing~\cite{SUPP}. 

The two lower bands in $\text{T}_4$ and $\text{T}_4'$ are stable with finite Bogoliubov gaps; the state evolution will follow the instantaneous eigenstate as $k(t)$ varies slowly, which exhibits a period of $4\pi/\epsilon$, exactly twice that of the Bloch Hamiltonian in Eq.~\ref{eq:Ham}. We identify this period-doubling as a discrete time crystal stabilized by nonlinear topology, 
thereby generalizing the concept of time crystalline order~\cite{Zhang2017ObservationDiscretea,Sacha2017TimeCrystals,RevModPhys.95.031001}. To probe the period-doubling, we can prepare an initial $k$-space wave packet in the instantaneous eigenstate and apply a weak gradient potential $\epsilon l$ in the synthetic lattice that is equivalent to incorporating the adiabatic parameter $k=k(0)+\epsilon t$ into the Hamiltonian~\cite{SUPP}. The evolution based on nonlinear dynamical equation is shown in Fig.~\ref{fig:period}b. In contrast to the ordinary {Bloch oscillation} observed for small-amplitude wave packet with $gn_k<|\delta J|$, the large-amplitude wave packet with peak nonlinearity $U=g\rho_0>J+|\delta J|$ exhibits a clear period-doubling, characterized by the relative phase $\varphi_k$ being restored only after a duration of $\epsilon t=4\pi$.  
Due to amplitude dropping, the dynamics becomes either nonadiabatic or ordinary {Bloch oscillations} in the {small} wings of the large-amplitude wave packet~\cite{SUPP}.

To validate our results, we also compute the full quantum eigennergy for each $k$ with fixed $gn_k$~\cite{SUPP}. Fig.~\ref{fig:period}c shows the full quantum bands and corresponding winding phase $\varphi_k=\arg[\langle \hat{a}^\dagger_k \hat{b}_k\rangle]$. The mean-field band crossing transforms into avoided crossings in the quantum regime, with gaps decreasing exponentially with photon number.
The behavior of the quantum ground state in the thermodynamic limit will coincide exactly with the predictions of the mean-field approximation. Even for finite photon numbers, the exponentially small gap can be much smaller than the applied gradient potential. Consequently, the {Bloch oscillation} dynamics is indistinguishable from those of true band crossings, and the period doubling persists in the quantum regime. However, the discrete time crystal reduces to a prethermal time crystal {with an exponentially long lifetime} due to the exponentially small {many-body} anti-crossing gaps~\cite{SUPP}. 

\begin{figure}[t]
    \centering
    \includegraphics[width=0.48\textwidth]{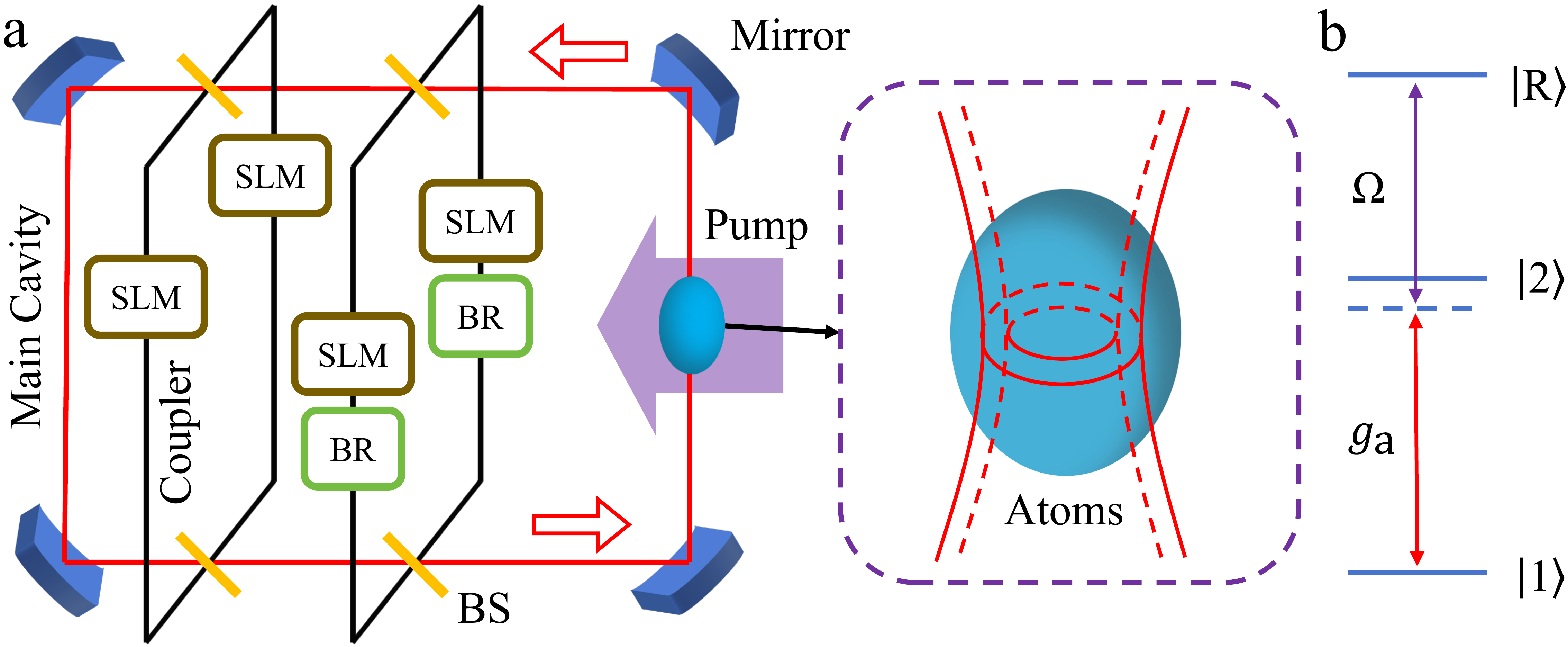}
	\caption{\justifying{\textbf{Schematic of the proposed experimental setup.} (a) OAM modes in the main cavity are coupled by coupler cavities consisting spatial light modulators (SLMs). The beam rotators (BRs) would induce a tunneling phase $e^{il\pi}=(-1)^l$~\cite{Luo2018TopologicalPhotonic}. (b) Nonlinearity is introduced by coupling the cavity photon with atomic Rydberg states through Raman process, with atom-cavity coupling $g_\text{a}$  and external pump $\Omega$.}}\label{fig:setups}
\end{figure}

\textit{\textcolor{blue}{Experimental consideration---}}Synthetic lattices engineered from atomic or photonic internal states have established themselves as a versatile platform for probing topological physics~\cite{An2017DirectObservation,An2021NonlinearDynamics,Celi2014SyntheticGaugea,Deng2022ObservingQuantum,Dutt2020SinglePhotonic,Gadway2015AtomopticsApproacha,Li2022AtomopticallySynthetic,Liang2024ChiralDynamicsa,Luo2015QuantumSimulationa,Luo2018TopologicalPhotonic,Lustig2019PhotonicTopological,Mancini2015ObservationChiral,Ozawa2016SyntheticDimensionsa,Ozawa2019TopologicalQuantum,Ren2023NonreciprocalDynamics,Stuhl2015VisualizingEdge,Wang2019SynthesizingArbitrary,Wang2024OnchipPhotonic,Yang2022SimulatingTopological,Yang2022TopologicalBand,Yu2025ComprehensiveReview,Yuan2018SyntheticDimensiona,Yuan2019PhotonicGauge,Cheng2025NonabelianLattice,Bouhiron2024RealizationAtomic,Yuan2020CreatingLocally,An2018CorrelatedDynamics,Wang2023TestingUniversality,Xie2020TopologicalQuantum,Chen2025InteractiondrivenBreakdown,Chen2024QuantumWalks,Chen2024StronglyInteracting,Kanungo2022RealizingTopological}. A distinctive strength of this approach lies in its inherent capacity to engineer long-range couplings (both in tunneling and interactions). Intriguing nonlinear dynamics have been revealed in atomic momentum lattices~\cite{An2018CorrelatedDynamics,Yuan2020CreatingLocally,Bouhiron2024RealizationAtomic,Wang2023TestingUniversality,Xie2020TopologicalQuantum,Chen2025InteractiondrivenBreakdown} and coupled synthetic Rydberg lattices~\cite{Chen2024QuantumWalks,Chen2024StronglyInteracting}.
We propose to implement our model using the OAM modes $e^{-il\theta}$ of photons inside a degenerate cavity~\cite{Luo2015QuantumSimulationa}, as shown in Fig.~\ref{fig:setups}a, where $l$ is the mode number with $\theta$ being the azimuthal angle. Two couplers, with SLMs changing the OAM states and BRs inducing the phase $(-1)^l$, generate the tunnelings of $J$ and $\delta J$ terms, respectively~\cite{SUPP}. To introduce nonlinearity, we couple the cavity photons with the Rydberg states of an atomic ensemble through a Raman process~\cite{clark2020observation,vsibalic2018rydberg,Georgakopoulos2018TheoryInteractinga,Pritchard2010CooperativeAtomlight}, as shown in Fig.~\ref{fig:setups}{b}. The hybridization of cavity photons with Rydberg excitations (i.e., formation of polaritons) would effectively introduce a photon-photon contact interaction in the $\theta$ space~\cite{SUPP}, which in the synthetic lattice space corresponds to all-to-all interactions preserving total OAM [Eq.~(\ref{eq:Hint})]. Here all the nonlocal density and exchange interactions are in resonance due to the degeneracy of OAM modes, which is fundamentally different from the atom-momentum based synthetic lattice~\cite{Wang2023TestingUniversality,An2018CorrelatedDynamics,Xie2020TopologicalQuantum,Chen2025InteractiondrivenBreakdown} where real-space contact interactions are reduced to local on-site interactions in the synthetic lattice due to energy mismatch. The gradient potential driving {Bloch oscillation} can be realized by inserting a beam rotator into the main cavity~\cite{SUPP}.

\begin{figure}[bt]
    \centering
    \includegraphics[width=0.48\textwidth]{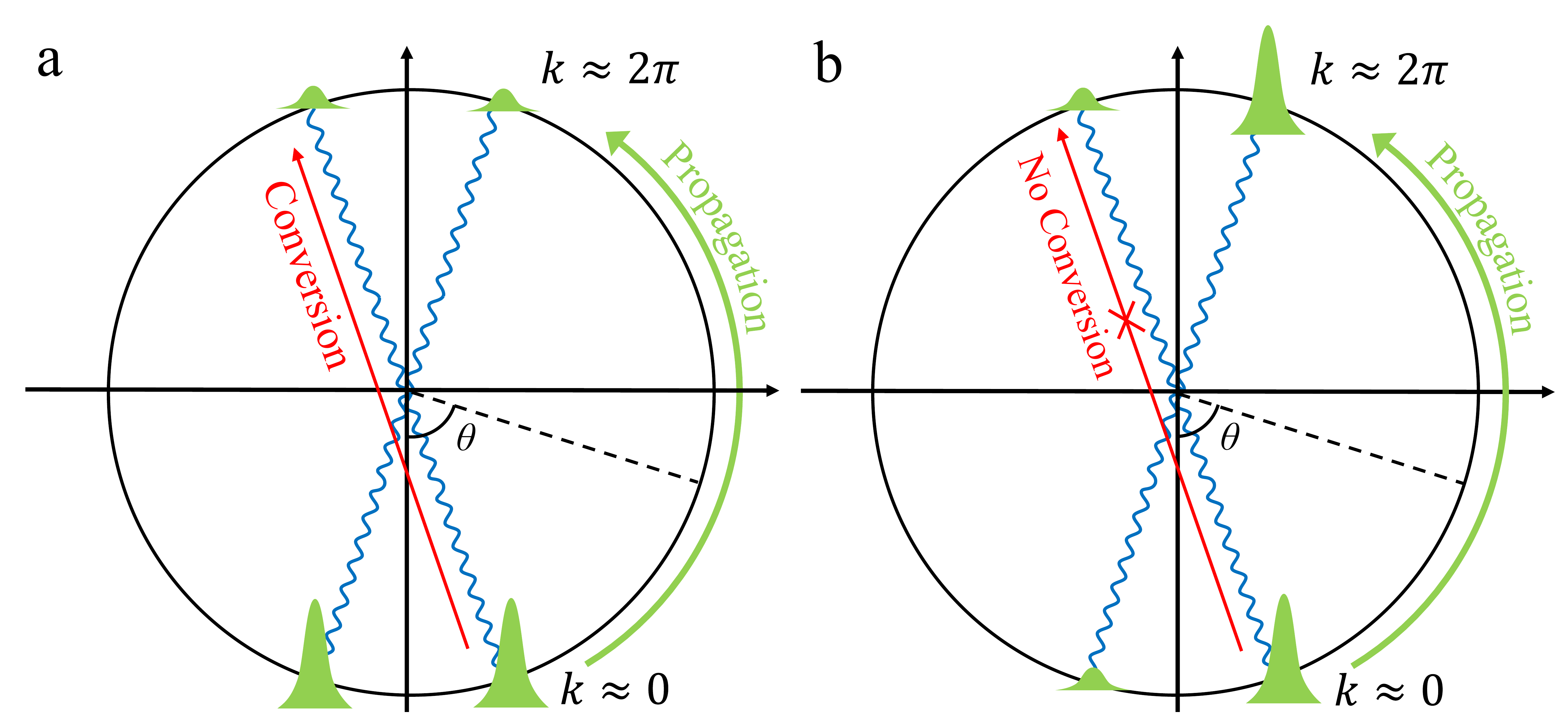}
	\caption{\justifying{\textbf{Adiabatic evolution in $\theta$-space.} (a) The wave packet eventually converts (red arrow) from $\theta$ to $\theta+\pi$ during the propagation (green arrow) for weak nonlinearity, restoring its initial state with a period of $\pi$.  (b) Strong nonlinearity suppresses the conversion, doubling the period to $2\pi$.
    }}\label{fig:wavepackets}
\end{figure}

The mechanism behind period doubling becomes more transparent in the $\theta$-space representation, where $\theta$ acts as the synthetic momentum space with reciprocal lattice vector $\pi$~\cite{SUPP}. In the linear regime, the eigenmode wave packet initialized at $\theta=0$ converts from $\theta$ to $\theta+\pi$ during propagation, restoring its initial state with a period of $\pi$ (see Fig.~\ref{fig:wavepackets}a). In contrast, strong nonlinearity suppresses this conversion, thereby doubling the period to $2\pi$ (see Fig.~\ref{fig:wavepackets}b). In the presence of photon loss, the decay of the wave packet amplitude during propagation weakens the effective nonlinearity. Nevertheless, the period-doubling response persists over the initial oscillation cycles for realistic low losses. Eventually, as the loss rate increases, this nonlinear effect diminishes, and the dynamics crosses over to ordinary {Bloch oscillations}~\cite{SUPP}. Moreover, although the interaction in Eq.~(\ref{eq:Hint}) naturally applies to our synthetic lattice system, the above {Bloch oscillation} dynamics remains robust when the interaction exhibits small deviations~\cite{SUPP}.

\textit{\textcolor{blue}{Conclusion and discussion}}---In summary, we have explored the interplay between topology and nonlocal nonlinearity in a synthetic SSH lattice. The unique all-to-all nonlinear coupling maintains an effective chiral symmetry that gives rise to a quantized nonlinear winding and Berry phase, as captured by our Bogoliubov nonlinear adiabatic theory. Increasing nonlinear strength drives a sequence of topological transitions, characterized by the emergence of swallowtail band structures and band swapping, along
with transitions of nonlinear edge states. The band swapping leads to fractional windings and multi-period {Bloch oscillations}, a discrete time crystalline behavior stabilized by nonlinear topology. Remarkably, strong nonlinearity can even induce an emergent topological phase where the single-particle Hamiltonian is topologically trivial. These nonlinear topological physics can be investigated experimentally using photons in a degenerate optical cavity with Rydberg-mediated interactions. 

Local real-space interactions usually project into a complex, nonlocal form in synthetic space, rendering them cumbersome to treat. The general impact of nonlocal nonlinearity on topological properties in synthetic lattices remains an open frontier. {For sufficiently strong interactions, quantum many-body physics beyond mean-field remains an important future direction.}
Our findings provide a unified framework for understanding unique nonlinear topological phases in photonic synthetic dimensions and open new avenues for realizing robust topological dynamics in synthetic quantum platforms.

\textit{Acknowledgments.---}This work was funded by the National Natural Science Foundation of China (Grants No. 12574544 and No. 12474366) and Innovation Program for Quantum Science and Technology (Grant No. 2021ZD0301200).
XWL also acknowledges the support from USTC start-up funding and thanks Chuanwei Zhang, Bryce Gadway and Junpeng Hou for stimulating discussions. HP is supported by the Welch Foundation (Grant No. C-1669) and the US NSF (PHYS-2513089).

%


\setcounter{figure}{0} \renewcommand{\thefigure}{S\arabic{figure}} %
\setcounter{equation}{0} \renewcommand{\theequation}{S\arabic{equation}}

\section*{Supplementary information}

\subsection{General Nonlinear Adiabatic Theory}
The total Hamiltonian $\mathcal{H}_\text{tot}=\mathcal{H}_0+\mathcal{H}_\text{int}$ in the Bloch momentum space is 
\begin{equation}
    \mathcal{H}_\text{tot}=
    \int dk \,\hat{\psi}_k^\dagger \begin{pmatrix}
        -g\hat{n}_k&\hat{h}^\dagger\\
        \hat{h}&-g\hat{n}_k
    \end{pmatrix} \hat{\psi}_k,
    \label{eq:Ham}
\end{equation}
where $\hat{\psi}_k=[\hat{a}_k,\hat{b}_k]^T$ with $\hat{a}_k=\frac{1}{\sqrt{2\pi}}\sum_n \hat{a}_n e^{-ink}$ (similar for $\hat{b}_k$), $\hat{h}=J_1+J_2e^{ik}-g\hat{a}^\dagger_k\hat{b}_k-g\hat{a}_k \hat{b}_k^\dagger e^{ik}$, and $\hat{n}_k=\hat{a}_k^\dagger \hat{a}_k+\hat{b}_k^\dagger \hat{b}_k$. 
It can be shown that the Boson field operators $\hat{a}_k$ and $\hat{b}_k$ satisfy the commutation relations 
\begin{eqnarray}
    [\hat{a}_k,\hat{a}_{k'}^\dagger]=[\hat{b}_k,\hat{b}_{k'}^\dagger]=\delta(k-k'). 
\end{eqnarray}
The Bloch momentum $k$ is a good quantum number in our system. For a given Bloch momentum $k$, the dynamics of the field operators can be obtained through the Heisenberg equations
\begin{eqnarray}
    i\partial_t \hat{a}_k=[\hat{a}_k,\mathcal{H}_{\rm tot}] \nonumber\\
    i\partial_t \hat{b}_k=[\hat{b}_k,\mathcal{H}_{\rm tot}]
\end{eqnarray}
According to commutation relations, it is straightforward to derive
\begin{eqnarray}
    i\partial_t\hat a_k=[\hat{a}_k,\mathcal{H}_{\rm tot}]
    =\begin{pmatrix}
        -2g\hat n_k&\hat h_\text{eff}^\dagger
    \end{pmatrix}\cdot\begin{pmatrix}
        \hat a_k\\
        \hat b_k
    \end{pmatrix},
\end{eqnarray}
where $\hat{h}_\text{eff}=J_1+J_2e^{ik}-2g\hat{a}_k^\dagger\hat{b}_k-2g\hat{a}_k \hat{b}^\dagger_k e^{ik}$. Similarly for 
$\hat b_k$.
Therefore, the dynamics of the field operator $\hat{\psi}_k=[\hat{a}_k,\hat{b}_k]^T$ is governed by
\begin{equation}
    i\partial_t \hat{\psi}_k=\hat{H}_\text{eff}(k)\hat{\psi}_k
\end{equation}
with effective Hamiltonian 
\begin{equation}
    \hat{H}_\text{eff}(k)=\begin{pmatrix}
        -2g\hat{n}_k& \hat{h}^\dagger_\text{eff}\\
        \hat{h}_\text{eff}&-2g\hat{n}_k
    \end{pmatrix}
\end{equation}

By replacing operators with mean-field c-numbers $\hat{\psi}_k\rightarrow\psi_k=[a_k,b_k]^T$, the above equation gives rise to the nonlinear Schr\"odinger equation (also known as the Gross-Pitaevskii equation).
The nonlinear eigenmodes, satisfying $H_{\rm eff}(k)\psi_{m,k}=E_m\psi_{m,k}$, are stationary solutions to 
the nonlinear Schr\"odinger equation, with $E_m$ the dynamical eigenenergy which corresponds to the chemical potential.
In the regime where the mean-field treatment is valid, these eigenmodes represent the coherent-state wavefunctions of the bosonic field (i.e., the optical field in photonic systems or Bose condensates in atomic systems).
The nonlinear eigenmodes are not the full quantum eigenstates of the many-body Hamiltonian $\mathcal{H}_{\rm tot}$. 

Generally, the mean-field approach is valid for weakly interacting bosonic systems, here ``weak" means the interaction strength between two individual particles (i.e., $g$) is much smaller than the kinetic energy (i.e., $J_1,J_2$), even though the effective nonlinearity $gn_k$ can become strong when the particle number is sufficiently large.
For photonic systems with nonlinear media, the dynamics is well captured by the mean-field nonlinear Schr\"odinger equation~\cite{Smirnova2020NonlinearTopological}, since the interaction is generally very weak at the single photon level. For cold atomic Bose gasses, the mean-field treatment works very well in ``weakly" interacting regime, where the scattering length is much smaller than the averaged separation between atoms~\cite{pitaevskii2016bose}.
When the interaction strength between two particles is comparable to their kinetic energy, the mean field treatment breaks down, the system should be described by strongly correlated many-body wavefunctions. For the physics studied in this work, the interaction strength is far below such a limit, and the mean-field treatment is valid. 
{We emphasize that, for sufficiently strong two-particle interactions $g$, the resulting many-body physics requires a full quantum treatment.}

Next, we adopt the mean-field treatment and develop the general nonlinear adiabatic theory. We begin by consider the above two-component nonlinear system. At fixed momentum $k$, our system is characterized by an  effective two-component nonlinear system $\psi=(a,b)^T$ (we omit the subscript $k$ here). First, we consider the instantaneous mean-field solution and the Bogoliubov excitation. In general, we have 
\begin{equation}
    H_\text{eff}=H_\text{eff}(\psi ,{\psi^*}).
\end{equation}
The corresponding Schr$\ddot{\text{o}}$dinger equation is
\begin{equation}
    i\frac{\partial}{\partial t} \psi =H_\text{eff}(\psi ,{\psi^*}) \psi ,\label{eq:eom}
\end{equation}
we define the  instantaneous eigenstate and the effective chemical potential as
\begin{equation}
    H_\text{eff}(\psi ,{\psi^*}) \psi_{m} =E_{m} \psi_{m} .
\end{equation}

{For the two-mode Bose condensate at fixed $k$, the global $U(1)$ phase fluctuation does not correspond to an observable excitation.}
Therefore, we focus on the gapped mode.
Then we expand the state using the Bogoliubov transform as in~\cite{pitaevskii2016bose,Baillie2017CollectiveExcitationsa}
\begin{equation}
     \psi(t) =e^{-iE_{m}t}[\sqrt{\rho_0}{\chi_{m}}+ \delta\psi(t) ],
\end{equation}
with the Bogoliubov excitation
\begin{equation}
     \delta\psi(t) = {\chi_{B}} [u {\alpha} e^{-i \omega t}+v^{*}{\alpha^*}e^{i \omega t}].
\end{equation}
Here $(u_k,v_k)$ are the instantaneous Bogoliubov modes with frequency $\omega_k$ and ${\alpha}$ is {fluctuation amplitude of the excitation}. The Bogoliubov mode has instantaneous spin state $|\chi_{B}\rangle$ orthogonal to the mean-field solution $|\chi_{m}\rangle$, {where $|\chi_{B}\rangle$ and $|\chi_{m}\rangle$ are the corresponding Dirac notations of the two spin-state vectors.} To solve for the Bogoliubov modes, we first substitute the state into the Schr$\ddot{\text{o}}$dinger equation~\eqref{eq:eom} and project onto $\langle\chi_B|$, keeping terms to the first order in $\delta \psi$
\begin{eqnarray}
    &&(E_m+\omega)u{\alpha} e^{i\omega t}+(E_m-\omega){\alpha^*} v^*e^{-i\omega t}v\nonumber\\
    &=&\langle\chi_B|D[H_{\text{eff}}]|\chi_m\rangle\left(u{\alpha} e^{-i\omega t}+v^*{\alpha^*} e^{i\omega t}\right)\nonumber\\
    &+&\langle\chi_B|D[H_{\text{eff}}]^*|\chi_m\rangle\left(u^*{\alpha^*} e^{i\omega t}+v{\alpha} e^{-i\omega t}\right)\nonumber\\
    &+&\langle\chi_B|H_{\text{eff}}|\chi_B\rangle\left(u{\alpha} e^{-i\omega t}+v^*{\alpha^*} e^{i\omega t}\right)
\end{eqnarray}
Here we define the displacement operator as:
\begin{equation}
    D[H_{\text{eff}}]=\sqrt{\rho_0}\sum_j\left.\frac{\partial H_\text{eff}(\psi ,{\psi^*})}{\partial \psi(j)}\right|_{\psi=\psi_m}\langle j|\chi_B\rangle,
\end{equation}
$\langle j|\chi_n\rangle$ and $\psi(j)$ denote the $j$-th components of $|\chi_n\rangle$ and $ \psi $.
For the effective Hamiltonian at momentum $k$, 
the displacement operator can be written as
\begin{equation}
    D[H_{\text{eff}}]=\begin{pmatrix}
        0&\delta h_{\text{eff}}^\prime\\
        \delta h_{\text{eff}}&0
    \end{pmatrix},
\end{equation}
where $\delta h_\text{eff}^\prime=-2U\langle1|\chi_B\rangle\langle\chi_m|2\rangle-2U\langle2|\chi_B\rangle\langle\chi_m|1\rangle e^{-ik}$ and ${\delta h}_\text{eff}=-2U\langle1|\chi_B\rangle\langle\chi_m|2\rangle e^{ik}-2U\langle2|\chi_B\rangle\langle\chi_m|1\rangle$ with $U=g\rho_0$. 

Then we reorganize the equations of the Bogoliubov modes according to the coefficients of ${\alpha}$ and ${\alpha^*}$; we have
\begin{eqnarray}
    A u+B u+\Delta^* v&=&\omega u+E_m u,\\
    A v+B v+ \Delta u&=&-\omega v+E_m v,
\end{eqnarray}
where
\begin{eqnarray}
    A&=&\langle\chi_B| H_\text{eff}(\psi ,{\psi^*})|\chi_B\rangle,\\
    B&=&\langle\chi_B| D[H_{\text{eff}}]|\chi_m\rangle,\\
    \Delta&=&\langle\chi_m|D[H_{\text{eff}}]|\chi_B\rangle=|\Delta|e^{i\phi_0}.
\end{eqnarray}
with $A,B$ real. So we get the Hartree-Fock BdG equations, which show the relationship between $u$ and $v$.
\begin{equation}
    \begin{pmatrix}
        \mathcal{L}-\omega&\Delta^*\\
        \Delta&\mathcal{L}+\omega
    \end{pmatrix}\begin{pmatrix}
        u\\
        v
    \end{pmatrix}=\begin{pmatrix}
        0\\
        0
    \end{pmatrix},\label{eq:BdG}
\end{equation}
where $\mathcal{L}=A+B-E_m$. From the BdG equations, the energy spectrum needs to satisfy
\begin{equation}
    \omega^2=\mathcal{L}^2-|\Delta|^2\label{spectrum}.
\end{equation}
Since $u$ and $v$ satisfy the bosonic commutation relation:
\begin{equation}
    |u|^2-|v|^2=1,
\end{equation}
we have
\begin{eqnarray}
    u&=&\frac{\Delta^*}{\sqrt{|\Delta|^2-(\mathcal{L}-\sqrt{|\mathcal{L}|^2-|\Delta|^2}})^2},\\
    v&=&-\frac{\mathcal{L}-\sqrt{|\mathcal{L}|^2-|\Delta|^2}}{\sqrt{|\Delta|^2-(\mathcal{L}-\sqrt{|\mathcal{L}|^2-|\Delta|^2}})^2}.
\end{eqnarray}
The two solutions of the BdG equation are not independent; we have considered the solution with $\omega>0$.

Now, we incorporate the adiabatic parameter $k=\epsilon t$ into the quantum states. The Hamiltonian and the states of the system will be related to the adiabatic parameter (i.e., time dependent). The system is characterized by a time-dependent two-level system with instantaneous Hamiltonian and solution
\begin{eqnarray}
    H_\text{eff}&=&H(\psi ,{\psi^*},k(t)),\\
     \psi_{m,k(t)} &=&\sqrt{\rho_0}{\chi_{m,k(t)}}.
\end{eqnarray}
Therefore, {for notational simplicity, we use the time parameter $t$ to label the adiabatic trajectory, so that the formalism applies to a general adiabatic parameter $k(t)$.}
During evolution, an initial nonlinear eigenstate becomes  
\begin{equation}
     \psi(t) =e^{i\gamma_m(t)-i\int^t E_{m}(t)dt'}[\sqrt{\rho_0}{\chi_{m}(t)}+ \delta\psi(t) ],\nonumber
\end{equation}
with Bogoliubov excitation
\begin{eqnarray}
     \delta\psi(t) &=& {\chi_{B}(t)} [u(t) \alpha(t) e^{-i\int^{t}\omega(t') dt'} \nonumber\\
        &+&v^{*}(t)\alpha^{*}(t)e^{i\int^{t}\omega(t') dt'}].
\end{eqnarray}
Again we substitute the time-dependent state into the equation of motion ~\eqref{eq:eom} to the first order in the adiabatic parameter $\epsilon$, and project onto $\langle\chi_B|$.
\begin{eqnarray}
    \operatorname{LHS}&=&i\sqrt{\rho_0}\langle\chi_B(t)|\partial_t|\chi_m(t)\rangle\nonumber\\
    &+&iu(t)\partial_t\alpha(t) e^{-i\int^{t}\omega(t') dt'}+i v^*(t)\partial_t\alpha^*(t) e^{i\int^{t}\omega(t') dt'}\nonumber\\
    &+&\left(E_m(t)+\omega(t)\right)u(t)\alpha(t) e^{-i\int^{t}\omega(t') dt'}\nonumber\\
    &+&\left(E_m(t)-\omega(t)\right)v^*(t)\alpha^*(t) e^{i\int^{t}\omega(t') dt'},\\
    \operatorname{RHS}&=&(A+B)u(t)\alpha(t) e^{-i\int^{t}\omega(t') dt'}\nonumber\\
    &+&(A+B)v^*(t)\alpha^*(t)e^{i\int^{t}\omega(t') dt'}\nonumber\\
    &+&\Delta^*u^*(t)\alpha^*(t)e^{i\int^{t}\omega(t') dt'}\nonumber\\
    &+&\Delta^*v(t)\alpha(t) e^{-i\int^{t}\omega(t') dt'}.
\end{eqnarray}
In an adiabatic process, the following quantity is of first-order in $\epsilon$:
\begin{equation}
    \frac{d\omega}{dt}\sim\frac{d|\chi_m\rangle}{dt}\sim\frac{d|\chi_B\rangle}{dt}\sim \frac{du}{dt} \sim \frac{dv}{dt}\sim\alpha\sim\frac{d\alpha}{dt}\nonumber\label{eq:order}.
\end{equation}
Recalling the solution of the BdG equations~\eqref{eq:BdG}, we find that the above equation becomes a general equation of the perturbation amplitude $\alpha(t)$ as
\begin{eqnarray}
    -\sqrt{\rho_0}C_1e^{i\phi_1}&=&u\frac{d}{dt}\alpha(t) e^{-i\int^{t}\omega(t') dt'}\nonumber\\
    &+&v^*\frac{d}{dt}\alpha^*(t) e^{i\int^{t}\omega(t') dt'},
\end{eqnarray}
where we have defined:
\begin{equation}\langle\chi_B|\partial_t|\chi_m\rangle=C_1e^{i\phi_1},
\end{equation}
with $C_1>0$.
The general solution of the perturbation amplitude $\alpha$ is given by
\begin{equation}
    \alpha(t)=\frac{i\sqrt{\rho_0}C_1}{\omega}(e^{i\phi_1}u^*-e^{-i\phi_1}v^*)e^{i\int^{t}\omega(t') dt'}\label{eq:solution}.
\end{equation}
The adiabatic condition reads $\alpha/\sqrt{\rho_0}\ll1$; that is
$\frac{\langle\chi_B|\partial_t|\chi_m\rangle}{\omega}\ll1$.

It is clear that this form of the perturbation amplitude satisfies the adiabatic requirement. Projecting the equation onto $\langle\chi_m|$, we arrive at the equation of motion for the geometric phase $\gamma_m(t)$
\begin{equation}
    \frac{d}{dt}\gamma_m(t)=i\langle\chi_m(t)|\partial_t|\chi_m(t)\rangle-A_{\text{nl}}
    \label{eq:gamma:SI}
\end{equation}
with a nonlinear geometric connection
\begin{eqnarray}
     A_{\text{nl}}(t)&=&\langle\chi_{m}|\sqrt{\rho_0}H^{(1)}_\text{eff}(\alpha)|\chi_{m}\rangle
     \label{eq:Anl:SI}
\end{eqnarray}
with $H^{(1)}_\text{eff}$ the first-order correction of $H_{\rm eff}$ that depends on $\alpha$,
\begin{equation}
    H^{(1)}_\text{eff}=\sum_j\left.\frac{\partial H_\text{eff}(\psi ,{\psi^*})}{\partial \psi(j)}\right|_{\psi=\psi_m}\delta\psi(j,t)+h.c..
    \label{eq:H1:SI}
\end{equation}
Following some straightforward derivations, and introducing the variable
\begin{equation}
    \langle\chi_m|D[H_{\text{eff}}(t)]|\chi_m\rangle=C_2e^{i\phi_2},
\end{equation}
with $C_2>0$,
we have 
\begin{eqnarray}
     A_{\text{nl}}(t)=2\operatorname{Re}\left[C_2\left[e^{i\phi_2}u+e^{-i\phi_2}v\right]\frac{\alpha(t)}{\sqrt{\rho_0}}e^{-i\int^{t}\omega(t') dt'}\right]\nonumber,
\end{eqnarray}
After substituting the solution of the excitation amplitude \eqref{eq:solution} into the nonlinear connection, we have
\begin{eqnarray}
    A_{\text{nl}}(t)&=&2\operatorname{Im}\left[\frac{C_1C_2}{\omega} (ue^{i\phi_2} + ve^{-i\phi_2})(u^*e^{i\phi_1}-v^*e^{-i\phi_1})\right]\nonumber\\
    &=&\frac{2C_1C_2}{\omega}(|u|^2+|v|^2)\sin(\phi_1+\phi_2)\nonumber\\
    & &-\frac{4C_1C_2}{\omega}|v|^2\frac{|\Delta|}{\mathcal{L}-\omega}\sin(\phi_1-\phi_2+\phi_0).
\end{eqnarray}
If we need $A_{\text{nl}}(t)=0$ at any time, we require that
\begin{eqnarray}
    (\phi_1-\phi_2+\phi_0)\mod\pi&=&0,\\
    (\phi_1+\phi_2)\mod\pi&=&0.
\end{eqnarray}
It is worth noting that the phases $\phi_1-\phi_2+\phi_0$ and $\phi_1+\phi_2$ are gauge independent.
These phases depend on the properties of the total Hamiltonian. 

For our system with effective chiral symmetry, the mean-field solution
reads $ \psi_m =\sqrt{\rho_0}|\chi_m\rangle$ with
\begin{equation}
    |\chi_m\rangle=\frac{1}{\sqrt{2}}[1,e^{i\varphi_m}].
\end{equation}
Where we have fixed the gauge of the mean-field solution by setting $a$ real. For the Bogoliubov excitation, we have 
\begin{equation}
    |\chi_B\rangle=\frac{e^{i{\phi_g}}}{\sqrt{2}}[1,-e^{i\varphi_m}],
\end{equation}
with phase ${\phi_g}$ being the gauge choice.
\newline
1) Regarding the phase $\phi_0$: The $\Delta$ term corresponds to the bosonic pairing (similar to the fermionic BCS pairing), the `pairing' phase $\phi_0$ can always be set to zero with proper gauge choice $\phi_g$. It can be derived that
\begin{equation}
    \Delta=U[1-\cos(2\varphi_m-k)]e^{i2\phi_g},
\end{equation}
leading to $|\Delta|=U[1-\cos(2\varphi_m-k)]$ and
$\phi_0=2\phi_g$. 
\newline
2) Regarding the phase $\phi_1$: The adiabatic process may excite the Bogoliubov mode, with $\phi_1$ being the coupling phase. The coupling between the mean-field solution with Bogoliubov mode is given by
\begin{equation}
    \langle\chi_B|\partial_t|\chi_m\rangle=C_1e^{i\phi_1}=\frac{1}{2} e^{-i(\phi_g+\frac{\pi}{2})} \dot\varphi_m.
\end{equation}
Therefore, we have the transition amplitude $C_1=\frac{\dot\varphi_m}{2}$ and the transition phase $\phi_1=-\phi_g-\frac{\pi}{2}$.
\newline
3) Regarding the phase $\phi_2$: This phase can be viewed as the projection phase of the displacement operator $D[H_{\rm eff}(t)]$ onto the eigenstate $|\chi_m\rangle$. For our chiral symmetric system, we find that
\begin{equation}
    \langle\chi_m|D[H_{\text{eff}}(t)]|\chi_m\rangle=U\sin(2\varphi_m-k)e^{i(\phi_g+\frac{\pi}{2})},
\end{equation}
which leads to $C_2=U\sin(2\varphi_m-k)$ and $\phi_2=\phi_g+\frac{\pi}{2}$.
Now it is straightforward to verify that our system satisfies $\phi_1-\phi_2+\phi_0=-\pi$, $\phi_1+\phi_2=0$.

We have focused on the two-component nonlinear system in the discussions above, now we extend our analysis to more general models. It is straightforward to generalize our nonlinear adiabatic Bogoliubov theory to a $q$-component condensate. Notice that the nonlinear Berry phase takes the same form as Eqs.~(\ref{eq:gamma:SI})-(\ref{eq:H1:SI}). The solution for $\alpha$ is central to our results. Now we have $q-1$ gapped Bogoliubov modes $\alpha_n$ with $n=1,2,\cdots,q-1$ having frequencies $\omega_n$. Also, the Bogoliubov modes $u\chi_B $ and $v\chi_B$ now take more general spinor forms
as $ u_n =[u_n(1),u_n(2),\cdots u_n(q)]^T$ and $ v_n =[v_n(1),v_n(2),\cdots v_n(q)]^T$ and they are orthogonal to $|\chi_m\rangle$.
We introduce the projection on the excitation space, and its $j$-th component reads
\begin{equation}
    \langle j|(I-|\chi_m\rangle\langle\chi_m|)\partial_t|\chi_m\rangle=C_1(j)e^{i\phi_1(j)}.
\end{equation}
Then we have the solution as
\begin{eqnarray}
    \alpha_n&=&\sum_{j=1}^{q}\frac{i\sqrt{\rho_0}C_1(j)}{\omega_n}[e^{i\phi_1(j)}u_n^*(j)\nonumber\\
    &-&e^{-i\phi_1(j)}v_n^*(j)]e^{i\int^{t}\omega_n(t') dt'}\label{eq:qsolution}.
\end{eqnarray}
With these solutions, the computation of the nonlinear Berry phase is straightforward.

\subsection{General Period-multiplexing}
We consider a general $q$-band model with modulated tunneling
$J+\delta Je^{i l2\pi/q}$. 
The full Hamiltonian with interaction reads
\begin{eqnarray}
    \mathcal{H}_{\rm tot}&=&\sum_l J\hat c^{\dagger}_{l+1}\hat c_l+\delta Je^{i l2\pi/q} \hat c^{\dagger}_{l+1}\hat c_l+h.c.\nonumber\\
    &-&\frac{g}{2\pi}\sum_{l_1,l_2,l_3,l_4} \delta_{l_1+l_2,l_3+l_4} \hat{c}_{l_1}^\dagger  \hat{c}_{l_2}^\dagger \hat{c}_{l_3} \hat{c}_{l_4}\label{eq:Ham:SI:q},
\end{eqnarray}
where $q$ is an integer and $c_l$ is the annihilation operator of a particle at the $l$-th site. Since there are $q$ sublattice sites in each unit cell, by representing the OAM mode using unit-cell and sublattice indices, we have:
\begin{eqnarray}
    \hat c_{q(n-1)+1}&\to& \hat a_{1,n},\nonumber\\
    \hat c_{q(n-1)+2}&\to& \hat a_{2,n},\nonumber\\
    &\vdots&\nonumber\\
    \hat c_{q(n-1)+q}&\to& \hat a_{q,n}.
\end{eqnarray}
The total Hamiltonian $\mathcal{H}_{\text{tot}}$ is
\begin{equation}
    \mathcal{H}_{\text{tot}}=\mathcal{H}_0+\mathcal{H}_{\text{int}}.\nonumber\\
\end{equation}
The first part is the single-particle Hamiltonian 
\begin{equation}
    \mathcal{H}_0=\int dk \hat\psi_k^\dagger \begin{pmatrix}
        0&J_1^*&\cdots&0&J_q e^{-i k}\\
        J_1&0&\cdots&0&0\\
        \vdots&\vdots&\ddots&\vdots&\vdots\\
        0&0&\cdots&0&J_{q-1}^*         \\
        J_q^* e^{i k}&0&\cdots&J_{q-1}&0
    \end{pmatrix}\hat\psi_k,\nonumber\\
\end{equation}
and the second one is the interaction part, which can be written as
\begin{equation}
    \mathcal{H}_{\text{int}}=\int \frac{dk}{q}\hat\psi_k^\dagger V\begin{pmatrix}
        H_{1}&&&\\
        &H_{2}&&\\
        &&\ddots&&\\
        &&&H_{q}
    \end{pmatrix}V^\dagger\hat\psi_k,\nonumber
\end{equation}
where
\begin{eqnarray}
    J_j&=&J+\delta Je^{i2\pi j/q},
    \\
    \left[V\right]_{mn}&=&e^{i[2\pi m(n-1)/q+m k/q-k]},\\
    H_{n}&=&\left|\left[V^\dagger\hat\psi_k\right]_{nn}\right|^2=\sum_{j,j'}\hat\psi_{j'}^\dagger V^\dagger_{j'n} V_{nj}\hat \psi_j,
\end{eqnarray}
where we define the multi-component spinor
\begin{equation}
    \hat\psi_k=[\hat a_{1,k},\hat a_{2,k},\cdots,\hat a_{q,k}]^\text{T}
\end{equation}
with $\hat a_{j,k}=\frac{1}{\sqrt{2\pi}}\sum_n \hat a_{j,n}e^{-ink}$. 
Setting $q=2$, we directly obtain our total Hamiltonian in the main text. Notice that $J_j=J_{q-j}^*$; therefore, the Hamiltonian preserves the inversion symmetry
$\mathcal{I} \mathcal{H}_k\mathcal{I}=\mathcal{H}_{-k}$ with 
$\mathcal{H}_k$ the Bloch Hamiltonian
(i.e., $\mathcal{H}_{\rm tot}=\int dk \mathcal{H}_k$), and the inversion operation is $\mathcal{I}: \hat{a}_{j,k}\leftrightarrow \hat{a}_{q+1-j,-k}$ (i.e., $\hat{a}_{j,n}\leftrightarrow \hat{a}_{q+1-j,-n}$).
This inversion symmetry ensures the quantization of the Berry phase in the linear limit. For vanishing nonlinearity $U=0$, we have
\begin{equation}
    \gamma=i\int_{-\pi}^{\pi} dk\langle\chi_k|\partial_k|\chi_k\rangle=(q-1)\pi
    \label{eq:gamma0:SI},
\end{equation}

As we discussed in the main text, fractional winding constitutes an intrinsic property of strong nonlinearity, regardless of whether the underlying linear regime is topological or trivial.
\begin{figure}[t]
    \centering
    \includegraphics[width=0.48\textwidth]{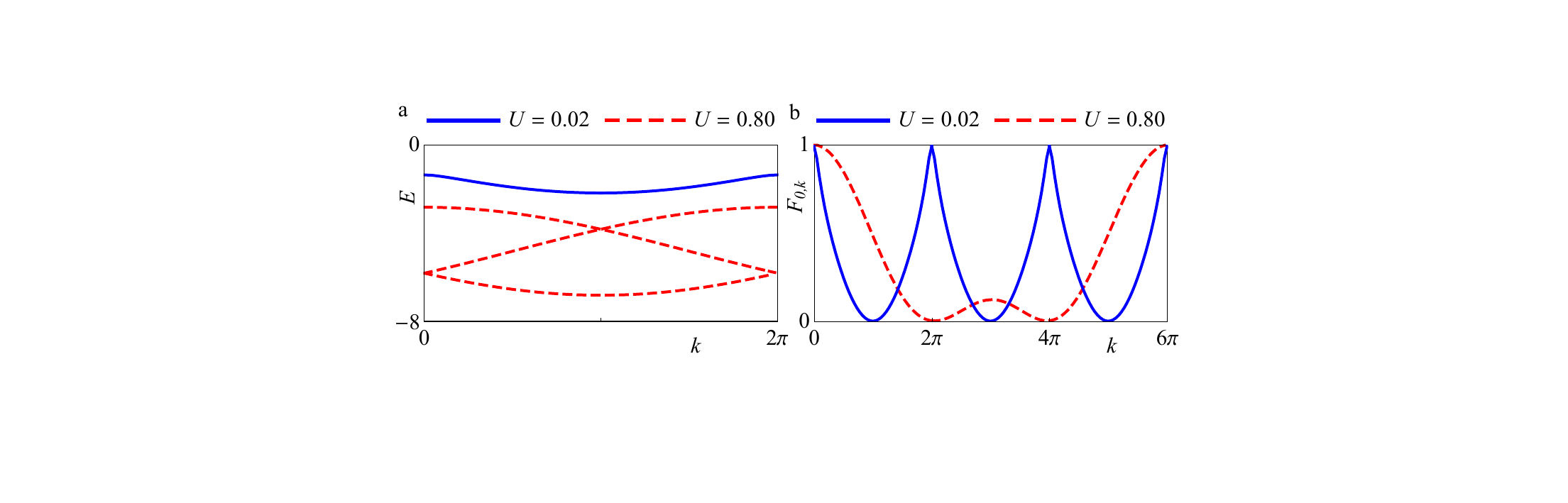}
    \caption{\justifying{\textbf{Period Tripling.} (a) Ground energy bands of the Hamiltonian Eq.~\ref{eq:Ham:SI:q} with $q=3$ under different nonlinearity. (b) The overlap of low-energy eigenstates between $k \neq 0$ and $k = 0$. Period-tripling appears in the strong nonlinear regime. Common parameters: $J=1$, $\delta J=0.3$.
    }}\label{fig:period:q3}
\end{figure}
The underlying principle allows for a direct generalization to period multiplexing. For example, a three-band model with $q=3$ exhibits a nonlinearity-induced period-tripling in its ground band; the corresponding band structure and state evolution are shown in Fig.~\ref{fig:period:q3}. For general $q$, the ground band evolves into $q$ intertwined bands in the strong nonlinear limit. The nonlinear eigenstate has a period of $q\times 2\pi$ in the Brillouin zone (BZ). These ground bands are stable with a well-defined nonlinear Berry phase. Though the full nonlinear Berry phase can be calculated using our Bogoliubov adiabatic theory, the nonlinear dynamical Berry connection $A_{\rm nl}$ is very complex for general $q$, making it difficult to determine whether the nonlinear Berry connection is zero or not. On the other hand, if we consider only the winding properties of the nonlinear eigenstates (i.e., the first part of the nonlinear Berry phase), we can introduce the phase winding of the ground energy sector as (we consider the strong nonlinear limit)
\begin{equation}
    \bar{\gamma}_q=i\int_{-q\pi}^{q\pi} dk\langle\chi_k|\partial_k|\chi_k\rangle
    \label{eq:gengamma:SI},
\end{equation}
where $|\chi_k\rangle$ is the unfolded eigenstate, which is defined continuously along $k$ and has a period of $2q\pi$. Because of the inversion symmetry $\mathcal{I}$, we have $|\chi_k\rangle=|\chi_{-k}\rangle$. Therefore, we must have ($\bar{\gamma}_q$ is defined up to a phase $2n\pi$ with integer $n$)
\begin{eqnarray}
    \bar{\gamma}_q=-\bar{\gamma}_q + 2n\pi
\end{eqnarray}
that is
\begin{eqnarray}
    \bar{\gamma}_q=n\pi.
\end{eqnarray}
Since $|\chi_k\rangle$ corresponds to a q-component spinor with spin $S=\frac{q-1}{2}$, the phase $\bar{\gamma}_q$ is closely related to the winding of the spin vector $(\langle S_x(k)\rangle,\langle S_y(k)\rangle,\langle S_z(k)\rangle)$. In the strong interaction limit, we find that
\begin{eqnarray}
    \bar{\gamma}_q=2S\pi=(q-1)\pi,
\end{eqnarray}
where $\langle S_z(k)\rangle\simeq0$ and $(\langle S_x(k)\rangle,\langle S_y(k)\rangle)$ winds around the origin once as $k$ varies from $-q\pi$ to $q\pi$.
Since $\bar{\gamma}_q$ is gauge invariant modulo $2\pi$, we have 
\begin{eqnarray}
\bar{\gamma}_q=\bigg\{\begin{array}{cc}
    0 & \text{if } q \text{ is odd} \\
    \pi & \text{if } q \text{ is even}
\end{array}
\end{eqnarray}
Since there are $q$ bands in the BZ,  each band contributes a fractional winding of
$2S\pi/q$ on average.

From the above discussion, we see that the period multiplexing  and fractional winding are general phenomena in our system. 
This period multiplexing manifests as a discrete time crystal stabilized by nonlinear topology.
We have focused on a simple tunneling modulation $\delta J e^{il2\pi/q}$. It would be interesting to extend our study to different types of tunneling modulations such as the generalized Aubry-Andre-Harper model. This may inspire some new research on symmetry properties and discrete time crystals in the future.

\begin{figure}[t]
    \centering
    \includegraphics[width=0.48\textwidth]{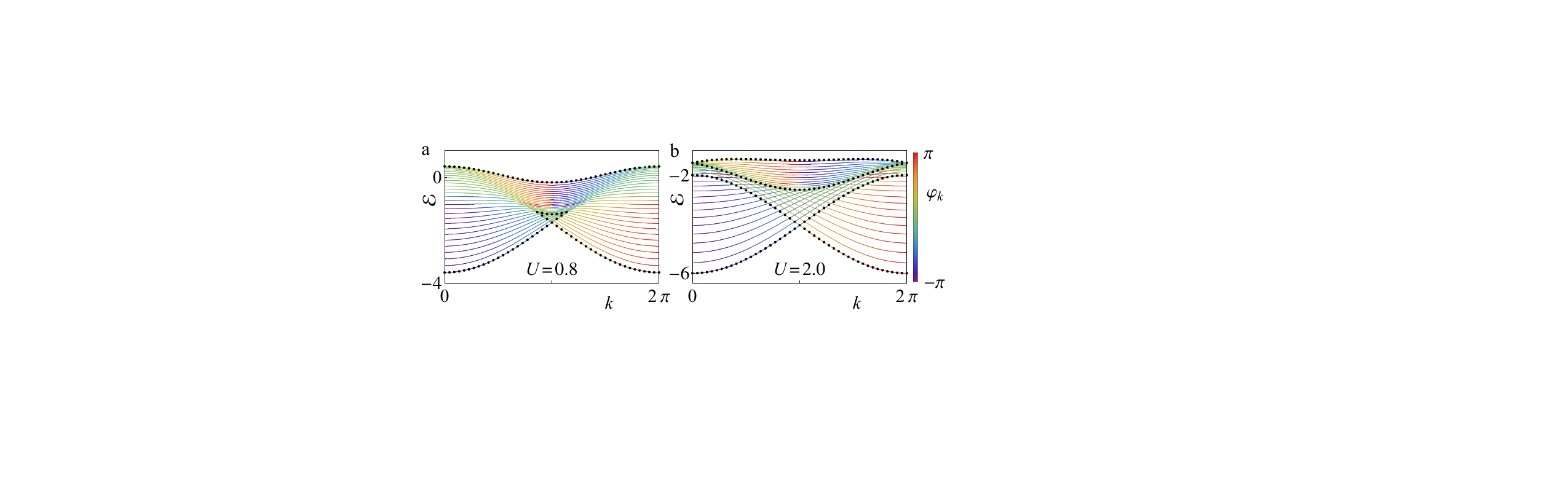}	\caption{\justifying{\textbf{Comparison of energy bands between full quantum and mean-field results.} (a, b) Energy bands obtained from the full quantum (colored solid lines) and the mean-field approximation (black dotted lines) at different interaction strengths. In both cases, the mean-field bands fully envelop the avoided crossings present in the full quantum results. Color bar indicates the phase $\varphi_k=\arg[\langle \hat{a}_k^\dagger \hat{b}_k\rangle]$. The total particle number of the system is $n_k=27$. We set $U=gn_k$, $J=1$ and $\delta J=0.3$.
    }}\label{fig:fullquantumband}
\end{figure}

\subsection{Full Quantum Energy Bands}
Since Bloch momentum is a good quantum number in our system, we consider the quantum Hamiltonian $\mathcal{H}_k$ at a given $k$. The full quantum energy bands can be obtained by directly diagonalizing the second-quantized two-mode model Hamiltonian
\begin{equation}
    \mathcal{H}_{k}= \hat{\psi}_k^\dagger \begin{pmatrix}
        -g\hat{n}_k&\hat{h}^\dagger\\
        \hat{h}&-g\hat{n}_k
    \end{pmatrix} \hat{\psi}_k,
\end{equation}

\begin{figure}[t]
    \centering
    \includegraphics[width=0.48\textwidth]{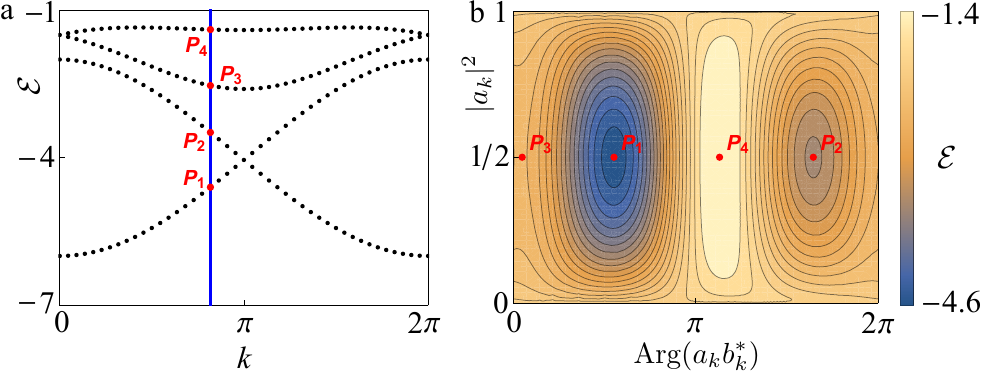}
    \caption{{\textbf{Mean-field energy and the contour map in the parameter space.}} (a) Mean-field energy per particle for each bands $\mathcal{E}_{m,k}=\langle \mathcal{H}_k\rangle_m/\rho_0$, with $\langle \mathcal{H}_k\rangle_m= \mathcal{H}_k(\hat\psi_k\rightarrow \psi_{m,k} )$. (b) Contour map of $\langle \mathcal{H}_k\rangle/\rho_0=\mathcal{H}_k(\hat\psi_k\rightarrow \psi_k )/\rho_0$ in the parameter space $ \psi_k =[a_k,b_k]^T$ at $k=2.57$ shown by the vertical blue line in (a). The nonlinear eigenmodes of the system correspond perfectly to the fixed points: the energy extrema (minima $P_1,P_2$; maximum $P_4$) and saddle point ($P_3$) on the energy landscape. Here we set $J=1$, $\delta J=0.3$, $U=g\rho_0=2.0$ same as that in Fig.~3c in the main text.}
    \label{fig:fixpoint}
\end{figure}

We note that $\hat n_k$ is a conserved quantity, it commutes with $\mathcal{H}_k$. Therefore, we can set a fixed total photon number $n_k$, and the relevant Hilbert space is $(n_k+1)$-dimensional, spanned by the Fock states $\{|\hat{a}_k^\dagger \hat{a}_k=n;\hat{b}_k^\dagger \hat{b}_k=n_k-n\rangle\}$ with $n=0,1,2,\cdots,n_k$. The nonlinear strength now becomes $U=gn_k$. Within this $(n_k+1)$-dimensional Hilbert space, we solve for the eigenenergy and the corresponding eigenstates through exact diagonalization
\begin{eqnarray}
    \mathcal{H}_k|\Phi_{m_Q,k}\rangle=n_k\mathcal{E}_{m_Q,k}|\Phi_{m_Q,k}\rangle,
\end{eqnarray}
with $m_Q=1,\cdots,n_k+1$ the quantum band index. The quantum eigenenergy per particle $\mathcal{E}_{m_Q,k}$ is plotted in Fig.~\ref{fig:fullquantumband}. As a comparison, the mean-field results are also shown in Fig.~\ref{fig:fullquantumband} (see the dotted lines). 
It is worth noting that the eigenenergy bands of $H_{\rm eff}$ shown in Fig.~1 in the main text correspond to the chemical potential, different from the true physical energy of the system. The mean-field energy should be the expectation of $\mathcal{H}_{k}$ (which is obtained by replacing the operators by mean-field solutions). In fact, the mean-field eigenmodes $\psi_{m,k}$ correspond to the extrema or saddle points of the mean energy $\langle\mathcal{H}_k\rangle=\mathcal{H}_k(\hat\psi_k\rightarrow \psi_k )$ in the parameter space $(a_k,b_k)$, as shown in Fig.~\ref{fig:fixpoint}. For example, the low-energy eigenmode minimizes $\langle\mathcal{H}_k\rangle$, and the system follows the corresponding solution adiabatically as $k$ varies in the Bloch oscillation. Furthermore, a constant correction $2g$ is applied to the mean-field energy $\mathcal{E}_{m,k}$, since when we use mean-field approximation by treating the operators into c-numbers, there will be a total $-2gN$ energy shift caused by commutation relation. Though the real energy bands take different shapes from the chemical potential bands, the appearance and merging of the swallowtail structures are similar for both of them, since they are related to the same nonlinear eigenmodes. Importantly, the winding properties of the quantum eigenstates are consistent with those obtained from the mean-field approximation, as shown by the color bar in Fig.~\ref{fig:fullquantumband}.

The quantized energy levels are bounded by the mean-field energies, and the mean-field energy levels envelop the net of anti-crossings in the quantized energy levels~\cite{Karkuszewski2002MeanFielda,Mulansky2011ImpurityBoseeinstein,Wu2006CommutabilitySemiclassicala}. The anti-crossing gaps decrease exponentially with photon number, so the behavior of the system in the thermodynamic limit will coincide exactly with the predictions of the mean-field approximation since the anti-crossings will become crossings.
Even for finite photon numbers (but large enough), 
the exponentially small gap could be much smaller than the weak gradient potential in the Bloch oscillation, where the anti-crossings act as crossings and the system will follow the mean-field solution, leading to period-doubling.
The mean-field predicted discrete time crystal is exact only in the thermodynamic limit with an infinite photon number, while for finite photon number, it becomes a prethermal time crystal due to the exponentially small anti-crossing gaps.
{In particular,
the nonlinear band crossing is exact and gives rise to an exact period-doubled Bloch oscillation at the mean-field level. At the full quantum level, however, the band crossing is expected to become an avoided crossing with an exponentially small gap that decreases with increasing photon number. As a result, each Bloch oscillation cycle induces only an exponentially small excitation probability, so that the period-doubled dynamics persists for an exponentially long time before eventual thermalization. In this sense, the exact discrete time crystal of the mean-field theory reduces to a prethermal time crystal at finite photon number.}

\begin{figure}[t]
    \centering
    \includegraphics[width=0.5\textwidth]{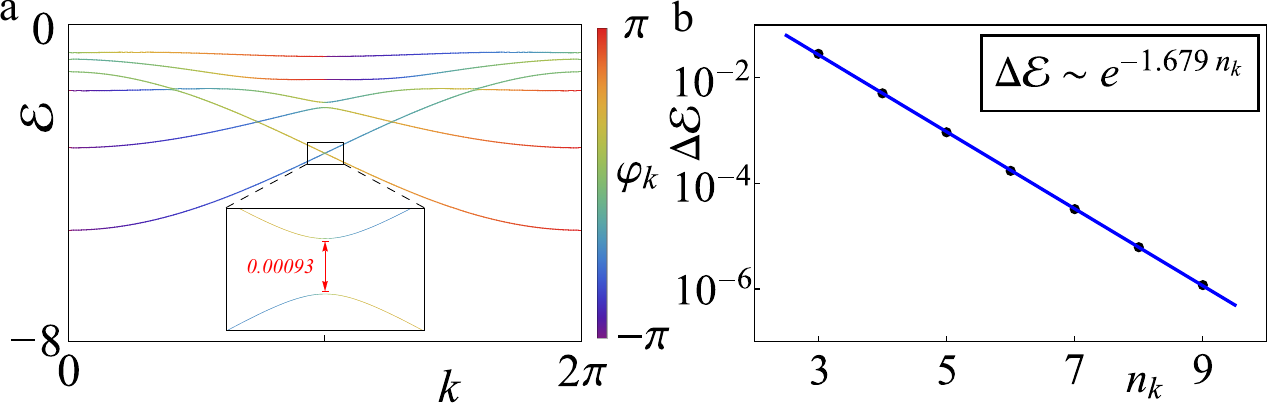}
    \caption{\textbf{Quantum energy gap as a function of $n_k$.} (a) Energy bands for $n_k=5$ {($\forall k$)}. (b) Energy gap at $k=\pi$ between the two lowest bands. As the number increases, the gap exhibits a clear exponential decay. A similar behavior is observed at all other avoided crossings within the fully quantum regime. We set $J=1$, $\delta J=0.3$ and $U=g{n_k}=2.0$.}
    \label{fig:energygap}
\end{figure}

To verify that the anti-crossing gap decreases exponentially with photon number, we investigate the quantum energy gap $\Delta \mathcal{E}$ between the ground state and the first excited state at $k=\pi$ for different photon numbers {while keeping $gn_k$ fixed}. As shown in Fig.~\ref{fig:energygap}, the gap is already very small even for $n_k=5$. As the photon number increases, the exponentially shrinking energy gap rapidly approaches zero. For $n_k=27$ the avoided crossing of the bands remains invisible even when significantly magnified. {For a given nonlinearity $g n_k$, the required two-photon interaction strength $g$ decreases as the photon number $n_k$ increases. In practice, $n_k$ is typically very large for coherent light fields, allowing strong nonlinear effects even for weak two-photon interactions, where the mean-field description remains valid.}

\subsection{Experimental consideration}
\subsection{Single-particle Hamiltonian}
The synthetic lattice model studied in this work can be realized using the OAM modes of photons inside a degenerate cavity, as shown in Fig.~4 in the main text. We consider the cavity modes with a narrow-ring shaped transverse density, so we can focus only on the dynamics along the azimuthal direction $\theta$ with the OAM mode profile $e^{-il\theta}/\sqrt{2\pi}$, where $l$ is the mode index. The $J$ tunneling term can be realized by a coupler cavity with spatial light modulators (SLMs), while the tunneling term $\delta J$ can be realized by a coupler cavity with SLMs and beam rotators (BRs)~\cite{Luo2018TopologicalPhotonic}. The SLM induces the change of OAM modes, while the BR rotates the beam by an angle $\theta_R$. 
The single-particle part of the Hamiltonian in Eq.~\ref{eq:Ham:SI:q} can be realized by setting $\theta_R=2\pi/q$. The gradient potential along the OAM synthetic dimension that drives Bloch oscillation can be realized by inserting a beam rotator into the main cavity, with the rotation angle $-\frac{\epsilon L}{2c}$ and cavity length $L$.

The tunneling parameters are governed by the reflection ratio of the beam splitters (BSs) and the optical path length of the coupler.
The photon with OAM $l$ can be reflected from the main cavity to the coupler by one BS, its OAM state is changed by $\pm1$ after passing through the SLM,
then the photon is reflected back into the main cavity by another BS. The total length of the coupler is designed for destructive interference. Therefore, the above process leads to tunneling of photons between the $l$-OAM mode and $(l\pm1)$-OAM modes in the main cavity $J e^{i\phi_J} \hat{c}_l^\dagger \hat{c}_{l+1}+h.c.$~\cite{Luo2018TopologicalPhotonic}.
The tunneling rate is $J\simeq \frac{rc}{L}$ with $r\ll 1$ the reflectivity  of the BS, $L$ the length of the main cavity, and $c$ the speed of light. The tunneling phase $\phi_J$ is determined by the phase-delay difference between the two arms of the coupler. For balanced arms without the BRs, the tunneling phase is zero, inserting the BRs can introduce an OAM dependent phase delay $e^{\pm il\theta_R}$ to the two arms, respectively. Since the two BRs rotate the OAM modes by an angle $\pm \theta_R$, which changes light field as $e^{il\theta}\rightarrow e^{il(\theta\pm\theta_R)}=e^{il\theta}e^{\pm il\theta_R}$, with $\theta$ the azimuth angle, leading to $\phi_J=l\theta_R$. As a result, one coupler realizes the tunneling 
$J \hat{c}_l^\dagger \hat{c}_{l+1}+h.c.$, the other coupler realizes the tunneling $\delta J e^{il\theta_R} \hat{c}_l^\dagger \hat{c}_{l+1}+h.c.$, our model corresponds to $\theta_R=2\pi/q$ with $q=2$ for the SSH model.
Finally, we discuss some details on how to 
realize the gradient potential along the OAM synthetic dimension that drives
Bloch oscillation. This can be done by inserting another BR into the main cavity with rotation angle $\theta_R=-\frac{\epsilon L}{2c}$, this BR leads to OAM dependent round-trip phase delay $\frac{\omega_l L}{c}-\frac{l\epsilon L}{2c}$ with $\omega_l$ the resonance frequency of OAM mode $l$. The resonance condition becomes OAM dependent $\frac{\omega_lL}{c}-\frac{l\epsilon L}{2c}=\frac{\omega_0L}{c}$, and thus $\omega_l=\omega_0+\epsilon l/2$, leading to the gradient potential $\delta_l=\omega_l-\omega_0=\epsilon l/2$.

Here, we show how the gradient potential affects the dynamical equation. 
We first omit the system Hamiltonian and focus on the dynamics solely under the gradient potential.
The corresponding Schr\"odinger equation:
\begin{equation}
    i\partial_t c_l(t)= \frac{\epsilon}{2} lc_l(t).
\end{equation}
with $c_l(t)$ the mean-field coherent state of the operator $\hat{c}_l$. In the unit cell and sublattice index, we have
\begin{eqnarray}
    i\partial_t\begin{pmatrix}
        a_n\\
        b_n
    \end{pmatrix}= n\epsilon\begin{pmatrix}
        a_n\\
        b_n
    \end{pmatrix}-\frac{\epsilon}{4}\sigma_z\begin{pmatrix}
        a_n\\
        b_n
    \end{pmatrix}-\frac{\epsilon}{4}\begin{pmatrix}
        a_n\\
        b_n
    \end{pmatrix}.
\end{eqnarray}
Transform to the Bloch momentum space, we obtain
\begin{eqnarray}
    i\partial_t\begin{pmatrix}
        a_k\\
        b_k
    \end{pmatrix}= -i\epsilon \partial_k \begin{pmatrix}
        a_k\\
        b_k
    \end{pmatrix}-\frac{\epsilon}{4}\sigma_z\begin{pmatrix}
        a_k\\
        b_k
    \end{pmatrix}.
\end{eqnarray}
We have dropped the constant energy shift $\epsilon/4$. Taking into account of the system Hamiltonian, the complete nonlinear Schr\"odinger equation reads
\begin{eqnarray}
    i\partial_t \psi(k,t) = \left[-i\epsilon \partial_k -\frac{\epsilon}{4}\sigma_z+H_{\rm eff}(k)\right]  \psi(k,t) .
\end{eqnarray}  
Note that there is an additional staggered potential (i.e. sub-lattice splitting) term $-\frac{\epsilon}{4}\sigma_z$. 
Although this term breaks the chiral symmetry and may render the winding number ill-defined as a topological invariant, its effect on dynamics becomes negligible in the weak-gradient limit. The only observable consequence of this term is a slight population imbalance between the two sublattices. Moreover, this term can be canceled by introducing an appropriate compensating staggered potential, we can omit the tiny $\sigma_z$ term in simulating the dynamics.
We use the characteristics method and parameterize the variables as
\begin{eqnarray}
    k&=&k_0+\epsilon\tau,\\
    t&=&\tau.
\end{eqnarray}
Then we have
\begin{equation}
     i\frac{d}{d\tau}\psi(k_0+\epsilon \tau,\tau)=H_\text{eff}(k_0+\epsilon \tau)\psi(k_0+\epsilon \tau,\tau).
\end{equation}
This equation is equivalent to incorporating the adiabatic parameter $k(t)=k_0+\epsilon t$ with $k(0)=k_0$ into the Hamiltonian. The dynamics can be obtained by solving this equation and transforming the results back into the original frame $(k,t)$.

\subsection{Nonlinear Rydberg Interaction}
\begin{figure}[t]
    \centering
    \includegraphics[width=0.48\textwidth]{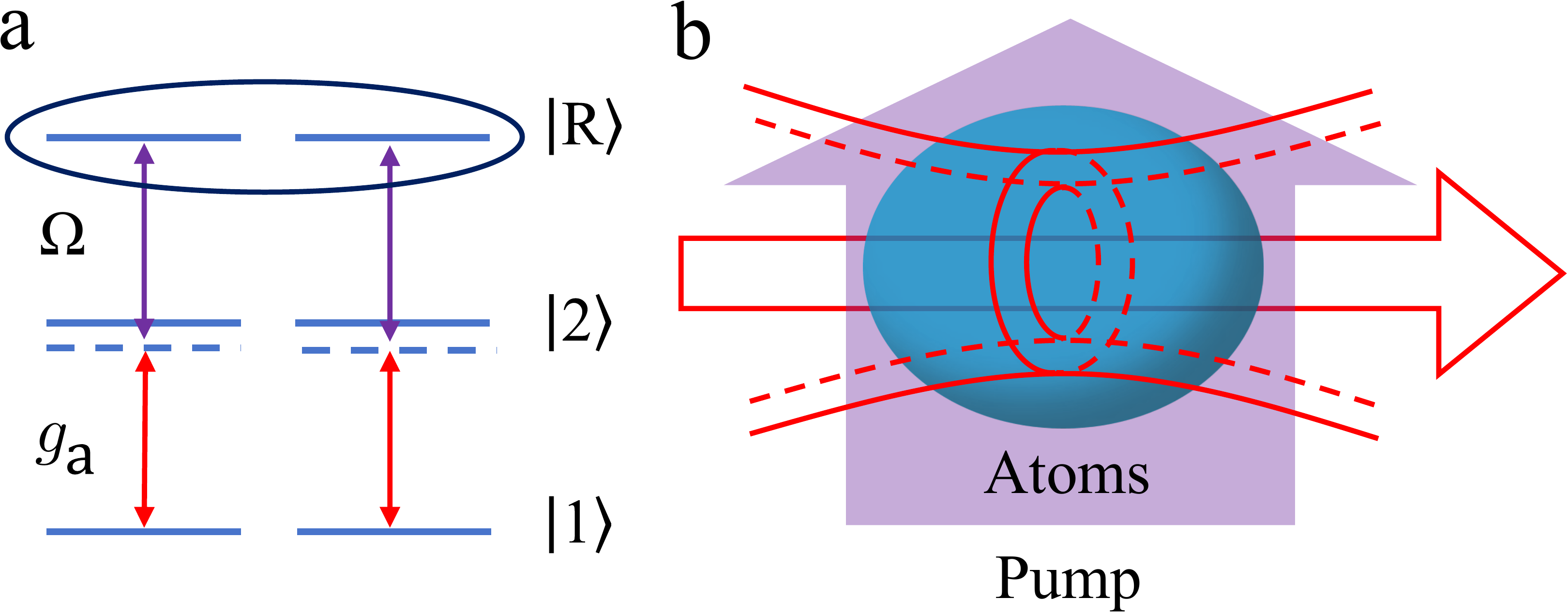}
	\caption{\justifying{\textbf{Photon interaction mediated by Rydberg atoms in a cavity.} (a) Energy level and Raman transition of the atoms. The ground state $|1\rangle$ is coupled to the excited state $|2\rangle$ through the cavity mode. The control pump beam then couples the excited state to the Rydberg state $|\text{R}\rangle$, leading to strong interactions. (b) The configuration of the cavity mode, pumping and the atom cloud.}}\label{fig:rydberg}
\end{figure}

To realize nonlinear interaction, we couple the cavity photons with the Rydberg states of an atomic ensemble through a Raman process, as shown in Fig.~\ref{fig:rydberg}. The hybridization of cavity photons with Rydberg excitations (i.e., formation of polaritons) would effectively introduce photon-photon interactions. 

Strong interactions between two atoms excited into Rydberg states originate from virtual photon exchange between them. This redistributes the atomic population between highly-excited states. Even two atoms in the same Rydberg state can interact by the Van der Waals interaction, which can be calculated in second-order perturbation theory and scales as
 $   \frac{n^8}{\Delta_E R^6}$,
where $n$ is known as the principal quantum number, $R$ is the inter-atomic distance and $\Delta_E$ is the energy defect, defined as
\begin{equation}
    \Delta_E\propto E(|n' P,n''P\rangle)-E(|nP,nP\rangle),
\end{equation}
where $P$ is the label used for the angular momentum state. Since for neighbouring states the energy defect is proportional to $n^{-3}$ when $n$ is large, the overall strength of these interactions scales as
\begin{equation}
    V(R)=-\frac{C_6}{R^6},
\end{equation}
where $C_6$ is a coefficient related to different Rydberg atoms~\cite{vsibalic2018rydberg,Georgakopoulos2018TheoryInteractinga}.
As shown in Fig.~\ref{fig:rydberg}, the coupling between cavity photons and the atom cloud will lead to effective photon-photon interactions through the formation of polaritons that hybridize with Rydberg states. We can express the polaritonic creation operators in the atomic basis as follows~\cite{Georgakopoulos2018TheoryInteractinga}
\begin{eqnarray}
    d_0^\dagger&=&\frac{\Omega}{\sqrt{g_{\rm a}^2(\mathbf{r})+\Omega^2}}c^\dagger(\mathbf{r})-\frac{g_{\rm{a}}(\mathbf{r})}{\sqrt{g_{\rm{a}}^2(\mathbf{r})+\Omega^2}}\phi^\dagger_{\text{R}}(\mathbf{r}),\nonumber\\
    d_{1,\pm}^\dagger&=&\frac{1}{\sqrt{2}}\left(\frac{g_{\rm{a}}(\mathbf{r})}{\sqrt{g_{\rm{a}}^2(\mathbf{r})+\Omega^2}}c^\dagger(\mathbf{r})\pm\phi^\dagger_2(\mathbf{r})\right.\nonumber\\
    &+&\left.\frac{\Omega}{\sqrt{g_{\rm{a}}^2(\mathbf{r})+\Omega^2}}\phi^\dagger_\text{R}(\mathbf{r})\right).
\end{eqnarray}
Here $\phi_\text{R}^\dagger(\mathbf{r})$ and $\phi_2^\dagger(\mathbf{r})$ are the bosonic creation operators for the Rydberg state and excited-state excitations of the ground-state atoms at  $\mathbf{r}=(r,\theta,z)$ from the center of the beam and $c^\dagger(\mathbf{r})$ is the bosonic creation operator for a photon of the cavity mode at the same place. Here $\Omega$ is the control field Rabi frequency and $g_{\rm{a}}(\mathbf{r})$ is the vacuum-Rabi coupling strength between a resonator photon localized at transverse location $z$ and a collective atomic excitation, and therefore must reflect the atom density. Indeed, it may be written as
\begin{equation}
    g_{\rm{a}}(\mathbf{r})\approx d_{12}\sqrt{\frac{L_\text{a}}{L}\frac{\rho_{\rm {a}}(\mathbf{r})\hbar\omega_{12}}{\varepsilon_0}},
\end{equation}
where $L_\text{a}$ is the length of the atomic ensemble along the resonator axis, $L$ is the length of the resonator itself, $d_{12}$ is the dipole moment of the atomic transition coupled to the optical resonator, $\omega_{12}$ is the angular frequency of this transition, and $\rho_{\rm {a}}(\mathbf{r})$ is the number density of atoms at location $\mathbf{r}$, which is uniform in our system.

If the interaction energy $V(\mathbf{r}-\mathbf{r'})=V(R)$ is small compared to the splitting between dark- and bright-polariton branches, the diagonal elements of the interaction Hamiltonian dominate, yielding the lowest-order polariton-projected effective interaction Hamiltonian~\cite{Georgakopoulos2018TheoryInteractinga}
\begin{eqnarray}
    H_{\text{int}}&=&\frac12\sin^4\frac{\theta_d}{2}\nonumber\\
    &\times&\left(\int d\mathbf{r} \int d\mathbf{r'}\ d_0^\dagger(\mathbf{r}) d_0^\dagger(\mathbf{r'}) V(\mathbf{r}-\mathbf{r'}) d_0(\mathbf{r'}) d_0(\mathbf{r})\right),\nonumber
\end{eqnarray}
where $\theta_d$ is defined as the dark state rotation angle. However, below a certain distance $R_B$ (for $V(R_B)=\Gamma$), known as the blockade radius~\cite{Pritchard2010CooperativeAtomlight}, the second atom is completely decoupled from the driving field, and any laser-induced dynamics bringing it to the given Rydberg state are blocked. Since the interaction decays rapidly with the distance $R$, in the region where the Rydberg blockade radius is smaller than the averaged inter-atomic distance, the interaction can be characterized by a contact interaction in real space. After integrating over $r$ and $z$, the Rydberg interaction reduces to
\begin{equation}
    H_{\text{int}}=-g\int d\theta\ d_0^\dagger(\theta) d_0^\dagger(\theta)d_0(\theta)d_0(\theta),
\end{equation}
with
\begin{equation}
    g\sim\frac{C_6}{R_B^5}\sin^4\frac{\theta_d}{2}.
\end{equation}
We consider the `nearly-photon' polaritons with small $\theta_d$ and approximate $d_0(\theta)$ by the photon operator $c(\theta)$; for simplicity, we rewrite the interaction as
\begin{equation}
    H_{\text{int}}=-g\int d\theta\ c^\dagger(\theta) c^\dagger(\theta)c(\theta)c(\theta).
\end{equation}
In the OAM basis,
we arrive at the nonlinearity presented in the main text
\begin{equation}
    \mathcal{H}_\text{int}=-\frac{g}{2\pi}\sum_{l_1,l_2,l_3,l_4} \delta_{l_1+l_2,l_3+l_4} \hat{c}_{l_1}^\dagger  \hat{c}_{l_2}^\dagger \hat{c}_{l_3} \hat{c}_{l_4}.
    \label{eq:Hint_supp}
\end{equation}
We can control the magnitude and the sign of the interaction strength by adjusting the type of Rydberg atoms, the atomic density and the pumping strength $\Omega$. It is worth noting that all the nonlocal density and exchange interactions here are in resonance due to the degeneracy of OAM modes (the resonance persists even under a gradient potential), which is fundamentally different from the atom-momentum-based synthetic lattice where real-space contact interactions are reduced to local on-site interactions in the synthetic momentum lattice due to energy mismatch of most terms.

Our synthetic lattice is constructed in the OAM space, the OAM $l$ and azimuthal angle $\theta$ are a pair of conjugate parameters (similar to the momentum and position). The interaction is generally local in the $\theta$ space (since it represents the real-space particle position) while along $l$ it conversely becomes long-range, rendering Eq.~(\ref{eq:Hint_supp}) a natural choice for the interaction term. The interaction between different Bloch momenta $k$ is difficult to introduce since it requires nonlocal interaction in real space; this is why previous studies have focused only on nonlinear topology with local interaction in real space lattices. Nevertheless, we could deliberately incorporate certain error terms to demonstrate the robustness of our results.

Errors in system interactions can be categorized into two types: the first is diagonal in $k$, which relates to a discrepancy in the interaction strengths; the second involves different Bloch momenta. 
1) For the first scenario: In general,  
such interaction errors may include inter- and intra-sublattice terms, taking the general form (other terms can be absorbed into the parameter $g$):
\begin{eqnarray}
    \mathcal{H}_k^{\rm error}
    &=&\hat{\psi}_k^\dagger\begin{pmatrix}
        \delta g \hat{n}_k&\delta \hat{h}^\dagger(k,\hat\psi_k)\\
        \delta\hat{h}(k,\hat\psi_k)&-\delta g\hat{n}_k
    \end{pmatrix}\hat\psi_k.
\end{eqnarray}
Here, the term $\delta\hat{h}$ preserves the chiral symmetry, it does not affect the topology and the phases of our system, though the phase boundaries may be slightly modified. The integer and fractional windings, as well as the swallowtail band structures, persist. However, the term $\delta g\hat{n}_k$ corresponds to interaction-induced staggered energy that breaks the chiral symmetry; consequently, the winding number is no longer a well-defined topological invariant for nonzero $\delta g$, and the mean-field eigenmodes have slightly unbalanced populations on the two sub-lattices.
This can be understood by noting that the topology of the SSH model originates from chiral symmetry protection. We find that for a sufficiently small symmetry-breaking term $\delta gn_k$, the swallowtail structure and band swapping (i.e. period-doubling Bloch oscillation) still emerge as we increase $U$. Also, the effect of $\delta g$ may be canceled out by introducing additional on-site potentials to the single-particle Hamiltonian.
2) For the second scenario: Interactions involving more than two Bloch momenta are unphysical in the context of synthetic dimensions, where the Bloch momentum plays the role of real-space position. Therefore, without loss of generality, we consider the interaction between $k$ and $k'$ taking the form of (analysis of other forms of errors is similar)
\begin{eqnarray}
    \mathcal{H}_{kk'}^{\rm error}=\delta g\cdot  \hat{a}_k^\dagger\hat{a}_k\cdot\hat{a}_{k'}^\dagger\hat{a}_{k'}.
\end{eqnarray}
We find that such interaction error does not affect the nonlinear eigenmodes.
The mean-field nonlinear dynamical equation now reads
\begin{eqnarray}
    i\partial_t \psi_k = H_{\rm eff}(k)  \psi_k +\delta g \begin{pmatrix}
        |a_{k'}|^2&0\\
        0&0
    \end{pmatrix}  \psi_k .
\end{eqnarray}  
We consider that the system occupies only a single Bloch momentum $k$:
\begin{eqnarray}
     \psi_k \neq 0;\quad \psi_{k'} =0
\end{eqnarray}  
Then the interaction error would have no effect on the system dynamics due to vanishing $a_{k'}$, and the nonlinear eigenmodes $ \psi_{m,k} $, the band structure and the topology remain unchanged. 
It can be derived that this interaction error gives rise to
Bogoliubov excitations at $k'$ with real energies $\frac{\delta g}{4}-E_{m,k}\pm\sqrt{\frac{\delta g^2}{16}+|J_1+J_2e^{ik'}|^2}$ ensuring the dynamical stability of the nonlinear eigenmodes $\psi_{m,k}$ at $k$, and these Bogoliubov gaps are also large and positive for the lower stable bands. Note that when $ \psi_{k'} $ is nonzero, the system is still dynamically stable if $\delta g$ is weak enough; however, this may induce weak effective on-site potentials for $ \psi_k $ that can break the chiral symmetry, whose effect has been discussed in point 1) above.

\subsection{$\theta$-Space Dynamics}
Similar to the position and momentum, here the azimuthal angle $\theta$ and OAM $l$ are a pair of conjugate parameters.
Our synthetic lattice space is represented by the OAM $l$, and thus the angle $\theta$ corresponds to the synthetic momentum
space. Since the lattice period along $l$ space is $q$, so the reciprocal lattice vector is $2\pi/q$ in the $\theta$ space.
We can write the Bloch wave function at $k$ as $\Psi_{m,k}(l)\equiv e^{ikn}\psi_{m,k}(l)$ in $l$-space, with unit cell index $n$ and band index $m$. Where $\psi_{m,k}(l)$ is the periodic part, we have
{$\psi_{m,k}(l=qn+j)=a_{j,m,k}$ with $j=0,1,\cdots,q-1$}. While in the $\theta$-space, we can write the Bloch wave function as $\Psi_{m,k}(\theta)$. Considering the reciprocal vector $2\pi/q$, we can conclude that $\Psi_{m,k}$ is non-zero only at azimuthal angle $\theta_k,\theta_k+\frac{2\pi}{q},\theta_k+\frac{4\pi}{q},\cdots$ with $\theta_k=k/q$. For the SSH model, the wave packet $\Psi_{m,k}$ has distributions only at $\theta_k$ and $\theta_k+\pi$. Therefore, the wave packet $\Psi_{m,k}$ consists of two wave packets at $\theta=\theta_k$ and $\theta=\theta_k+\pi$, as illustrated in Fig.~5 in the main text. As $k$ changes from $0$ to $2\pi$ during the Bloch oscillation, $\theta_k$ changes from $0$ to $\pi$.

To measure the Bloch oscillation, we initialize the wave packet at $k\simeq0$, the eigenmode $\Psi_{m,k}(\theta)$ consists of two wave packets at $\theta\simeq0$ and $\theta\simeq\pi$ with dominant distributions at $\theta\simeq0$, as illustrated in Fig.~5 in the main text, with a much larger wave packet at $\theta\simeq0$ than that at $\theta\simeq\pi$. (i) In the weak nonlinear regime with $U<\delta J$, the wave packet eventually converts from $\theta=\theta_k$ to $\theta=\theta_k+\pi$ as $\theta_k$ changes from $0$ to $\pi$ during one Bloch oscillation period,
thus the dominant wave packet returns to the initial position $\theta=2\pi$ at the end. 
As illustrated in Fig.~5(a), the green arrow indicates the propagation direction during the Bloch oscillation, while the red arrow indicates the conversion direction.  (ii) In the strong nonlinear regime with $U>J+|\delta J|$, the eigenmode $\psi_{m,k}$ returns to its initial state after $k$ traverses the Brillouin zone twice.
As $k$ changes from $0$ to $2\pi$, instead of returning to the initial position, the dominant wave packet in $\theta$-space propagates from $\theta=0$ to $\theta=\pi$ without wave-packet conversion, 
as illustrated in Fig.~5(b) in the main text. The nonlinearity prevents the wave-packet conversion, since it is local in $\theta$ space. The  wave packet returns to its initial position after two Bloch oscillation periods (i.e., when $k$ changes from $0$ to $4\pi$).

{The Bloch oscillation dynamics can be viewed as a smooth evolution of the Bloch momentum, $k(t)=k(0)+\epsilon t$. The $k$-space wave packet propagates along the Bloch momentum and the total density profile preserves its shape during the evolution. Therefore, the $k$-space wave packet does not split. The different Bloch oscillation periods originate from the evolution of the relative phase $\arg[a_k^*b_k]$ as $k$ traverses the Brillouin zone. Specifically, $\arg[a_k^*b_k]$ changes by $\pi$ for the double-period Bloch oscillation and by $2\pi$ for the ordinary Bloch oscillation. For a large-amplitude wave packet, the wings and center undergo different phase evolutions due to the reduction of the nonlinearity toward the wave-packet wings, with a crossover region where the dynamics becomes nonadiabatic. The different evolutions of the internal spinor structure at different values of $k$ can lead to distortions of the wave packet in the synthetic lattice space $l$. Since the wings contain only a small fraction of the total wave-packet weight, the overall dynamics is nevertheless dominated by the coherent double-period Bloch oscillation of the wave-packet center, which exhibits the characteristic $\pi$ phase shift between the $a$ and $b$ sublattices after one Bloch oscillation period.}

{The situation is different in the conjugate synthetic coordinate space $\theta$. For an initial state centered around $k\simeq 0$, the wave packet is predominantly localized near $\theta\simeq 0$. During the evolution, the central part of the wave packet, which exhibits double-period Bloch oscillations, evolves from $\theta\simeq0$ to $\theta\simeq\pi$ after one Bloch oscillation period, whereas the wings, which exhibit ordinary Bloch oscillations, return to their initial position. Consequently, the wave packet becomes spatially separated by approximately $\pi$ in the $\theta$ space. Since the reciprocal lattice vector associated with the synthetic lattice in the $\theta$ coordinate is $\pi$, this spatial separation does not manifest itself as a splitting of the wave packet in the Bloch momentum space $k$.}

\subsection{Experimental Parameter Estimation}
We consider a typical cavity length $L\approx0.3$ m. And the corresponding free spectral range is $\Omega_{\rm FSR}\simeq 2\pi\times 1$ GHz. We can choose the reflectivity of the beam splitter to be $r\sim0.12$, such that the tunneling $J,\delta J \sim r \Omega_{\rm FSR}/4\pi$ are of the order of $2\pi\times 10$ MHz~\cite{Luo2018TopologicalPhotonic}. The interaction between two Rydberg atoms can be up to the order of MHz for a typical atom distance of a few $\mu$m~\cite{vsibalic2018rydberg}. Although our polariton is nearly-photonic, with increasing photon number, one can easily reach the strong nonlinear region. 

The fractional winding and period doubling can be probed according to the Bloch oscillation, where the applied gradient potential should be weak enough to ensure that the system adiabatically follows the nonlinear eigenstates.
\begin{figure}[t]
    \centering
    \includegraphics[width=0.48\textwidth]{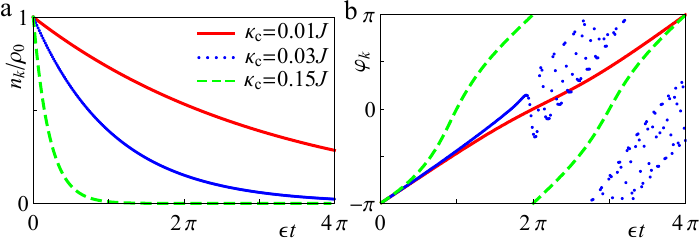}	\caption{\justifying{\textbf{Evolutions of total density $n_k/\rho_0$  and relative phase $\varphi_k=\arg[a^*_kb_k]$.}  Evolution of the total density (a) and the relative phase (b) of a wave packet with a strong initial interaction of $U=g\rho_0=10J$. $|a_k|\simeq|b_k|$ during the whole evolution. For $\kappa_{\text{c}}=0.01J$, period-doubling response persists over the initial oscillation cycles (with period $4\pi/\epsilon$). As the loss rate increases, this nonlinear effect diminishes, and the dynamics cross over to ordinary Bloch oscillations. We set $J=1$ and $\delta J=0.3$.}}\label{fig:loss}
\end{figure}
On the other hand, a realistic cavity has a linewidth $\kappa_\text{c}$ (i.e., the cavity decay rate) that determines the lifetime of cavity photons. The effect of Rydberg state decay (typically of the order of $10$ kHz) is negligible since the polariton is nearly photonic. Since the period doubling is induced by strong nonlinearity, the photon loss, which effectively decreases the nonlinearity, may destroy the period-doubling dynamics. Therefore, the gradient potential should be strong enough so that the Bloch oscillation period is shorter than the lifetime of the cavity photon. Here we consider $\epsilon=0.1 J$, which is small enough to ensure adiabaticity. In this case, we find that a realistic linewidth $\kappa_\text{c}\simeq 0.01 J\simeq 0.1$ MHz is small enough to ensure that the period-doubling response persists over the initial oscillation cycles, with an initial strong nonlinearity $U=10J$. The linewidth can be improved to $\kappa_c\lesssim0.001J\sim$ 10kHz~\cite{PhysRevLett.121.220405} by using high-performance optical elements, where period-doubling dynamics can last much longer. When the loss rate increases, this nonlinear effect diminishes, and the dynamics cross over to ordinary Bloch oscillations as expected. The numerical simulations are presented in Fig.~\ref{fig:loss}.

\subsection{Nonlinear Bulk-edge Correspondence}
As we discussed in the main text, if the topologically irrelevant nonlinear shift $gn_k$ is omitted from the Hamiltonian, the nonlinearity reduces to a purely inter-sublattice form. 
{Since the linear edge states are eigenstates of the chiral symmetry operator and are localized exclusively on a single sublattice, the nonlinear interaction vanishes for such states.}
Consequently, the edge states remain {identical to those in the linear limit, and
the correspondence between these edge states and the linear bulk topology is preserved irrespective of the nonlinear strength.}

However, in realistic settings, an ideal boundary is difficult to realize because nonlocal interactions inherently couple to OAM states beyond any single-particle boundary. Furthermore, the $k$-dependent density of the edge state prevents $gn_k$ from being omitted as a constant. 
\begin{figure}[t]
    \centering
    \includegraphics[width=0.48\textwidth]{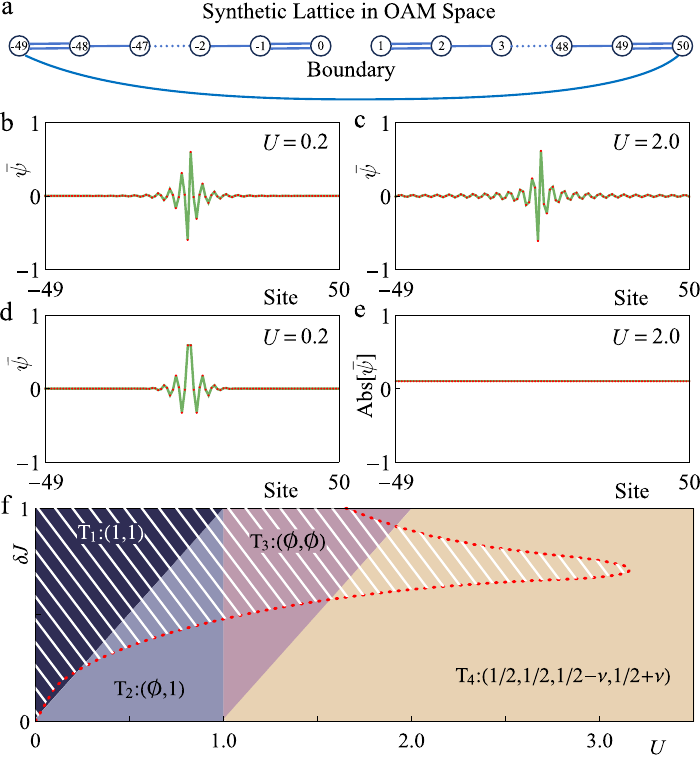}
	\caption{\justifying{\textbf{Nonlinear Edge States.} (a) The orbital angular momentum (OAM) lattice with an open boundary between site $n=0$ and $n=1$. (b--c) The anti-symmetric edge state persists under different nonlinear strengths.  (d--e) The symmetric edge state exists in the weak nonlinear region, but vanishes for strong nonlinearity where it evolves into a bulk state during iteration. The symmetric edge state vanishes beyond the boundary delineated by the red-dotted lines in (f). The edge states are purely real.  We have set $U=gN_{\text{edge}}$, $J=1$, $\delta J=0.3$ and $\bar\psi=\psi/\sqrt{N_{\text{edge}}}$.
    }}\label{fig:edgestates}
\end{figure}
Here, we consider opening a single-particle boundary at site $n=0$ by turning off the tunneling between $l=0$ and $l=1$, the interaction can still couple OAM modes from different sides of the boundary. Such a boundary can be constructed by introducing pinholes in beam splitters that connect the main cavity with the coupler cavity. Notice that we have assumed an identical narrow ring-shaped profile for all OAM modes at the SLM and the atomic cloud (such a requirement can be achieved by the degenerate cavity design), so we can have well-defined nonlinearity along the azimuthal angle $\theta$. On the other hand, near the beam splitter, different OAM modes can have different transverse profiles due to diffraction. Generally, the $l=0$ OAM state can have a much smaller beam spot compared to $l\neq 0$ states, and thus the pinhole beam splitter would not couple the $l=0$ mode, leading to the single-particle boundary, as shown in Fig.~\ref{fig:edgestates}a. 
We consider a periodic boundary on the large OAM ends to simulate the infinite possible OAM modes.

To numerically solve for the nonlinear edge states of the system with an open boundary at site $n=0$, we adopt an iterative approach.
Since the nonlinear solution is amplitude-dependent, we use the normalization $\langle \psi_{\rm edge}|\psi_{\rm edge}\rangle=N_{\rm edge}$, and define $U=gN_{\rm edge}$. 
We define the state-dependent effective mean-field Hamiltonian $H_{\rm eff}(\psi)$ in the OAM-lattice space, which can be obtained from the Heisenberg equation 
$i\dot{\hat{c}}_l=[\hat{c}_{l},\mathcal{H}_{\rm tot}]
$. After replacing the operators with mean-field c-numbers we arrive at the OAM-space nonlinear Schr\"odinger equation $i\dot{c}_l=\sum_{l'}[H_{\rm eff}(\psi)]_{ll'} \cdot c_{l'}$.  For the first step $s=1$, we can choose the initial state as the edge mode solution in the linear limit, or as a localized state at the boundary $|\psi_{s=1,\rm edge}\rangle=\sqrt{\frac{N_{\rm edge}}{2}}[0,\cdots, 0,b_{0}=\pm1,a_1=1,0,\cdots,0]^T$.
The iteration are as follows:
\begin{enumerate}  
    \item In the $s$-th step, we calculate the state-dependent effective Hamiltonian $H_{\rm eff}(\psi_{s,\rm edge})$ in the OAM space,
    and solve for its eigenstates $|\bar\psi_{s,j}\rangle=[\cdots,c_l^{(j)},c^{(j)}_{l+1},\cdots]^T$.
    Identify the $j_s$-th eigenstate $|\bar\psi_{s,j_s}\rangle$ that minimizes $\left||\bar\psi_{s,j}\rangle-|\psi_{s,\rm edge}\rangle\right|$.
    \item During the initial iterations $s<5$, the new edge states are obtained by $|\psi_{s+1,\rm edge}\rangle\propto (1-f_s)|\psi_{s,\rm edge}\rangle+f_s|\bar\psi_{s,j_s}\rangle$ with $f_s$ the Barzilai-Borwein dynamical relaxation factor.
    \item When $s\geq5$, we calculate the new edge state $|\psi_{s+1,\rm edge}\rangle$ from states $|\bar\psi_{s',j_s}\rangle$ and $|\psi_{s',\rm edge}\rangle$ with $s-5<s'\leq s$ according to the Anderson acceleration iteration method with the 5 latest steps.
    \item Repeat the iteration until $\left||\psi_{s+1,\rm edge}\rangle-|\psi_{s,\rm edge}\rangle\right|$ is less than a specified accuracy (here we use $10^{-10}$).  
\end{enumerate}

The results are shown in Fig.~\ref{fig:edgestates}b--e.
We find that for $J=1$, $\delta J>0$, there are two nonlinear edge-state solutions in the weak nonlinear region, but there is only one nonlinear edge-state solution (the anti-symmetric one) in the strong nonlinear region, as delineated by the red-dotted lines in Fig.~\ref{fig:edgestates}f. 
This is because, in addition to long-range couplings, the nonlocal interaction also induces effective nearest-neighbor tunneling $-\frac{g}{2\pi}c_0^*c_1 \hat{c}^{\dagger}_{l+1}\hat{c}_l+h.c.$ with amplitude $-\frac{g}{2\pi}c_0^*c_1$. For the symmetric edge state with attractive interaction, we have $-\frac{g}{2\pi}c_0^*c_1<0$, and thus nonlocal nonlinearity effectively weakens the nearest-neighbor tunneling $J_{1,2}$ (since we have chosen $J_{1,2}>0$), the effect of nonlinear long-range coupling becomes more prominent which delocalize the edge state in the strong nonlinear regime. As we increase $\delta J$ from 0, the linear edge state becomes more localized and requires a stronger nonlinearity to delocalize the symmetric edge state. On the other hand, for $\delta J\sim1$, we have $J_2\sim 0$, then
the interaction induced nearest-neighbor tunneling $-\frac{g}{2\pi}c_0^*c_1$ is dominant over $J_2$ and the edge-state solution becomes less localized, a weaker nonlinearity is enough to delocalize the symmetric edge state. Therefore, we have a boundary shown by the red-dotted lines in Fig.~\ref{fig:edgestates}f.
In contrast, for the antisymmetric edge state with $-\frac{g}{2\pi}c_0^*c_1>0$, the nonlinearity enhances $J_{1,2}$, thereby stabilizing the state which persists in the strong nonlinear regime.
We have verified the above discussion by changing the sign of $J_{1,2}$, and find that the antisymmetric edge state becomes the unstable one.

For $J=1$, $\delta J<0$ (i.e., $J_1>J_2$), starting from the localized state given above, we always end up with a bulk state in the iteration for arbitrary nonlinearity, implying the absence of nonlinear edge states. We emphasize that, though the band topologies in both phases $T_4$ and $T'_4$ are dominated by strong nonlinearity, the appearance of edge state is full determined by the staggered single-particle tunneling, since the nonlocal nonlinearity is uniform along the synthetic lattice.

\end{document}